\documentclass[final,3p,times,nopreprintline]{elsarticle}
\usepackage[utf8]{inputenc}
\usepackage[T1]{fontenc}
\usepackage{amsmath,amssymb,bm}
\usepackage{mathtools}
\usepackage{booktabs}
\usepackage{graphicx}
\usepackage{subcaption}
\usepackage{float}
\usepackage{hyperref}
\usepackage{cleveref}
\usepackage{algorithm}
\usepackage{algpseudocode}
\usepackage{xcolor}
\usepackage{siunitx}
\usepackage{tikz}
\usepackage{verbatim}
\usetikzlibrary{arrows.meta,positioning,calc,shapes.geometric,fit,backgrounds,decorations.pathreplacing,patterns}
\usepackage{enumitem}
\usepackage{microtype}
\newcommand{\grad}{\nabla}
\newcommand{\divop}{\nabla\!\cdot}

\newcommand{\vect}[1]{\bm{#1}}
\newcommand{\mat}[1]{\bm{#1}}


\crefname{equation}{Eq.}{Eqs.}
\Crefname{equation}{Eq.}{Eqs.}
\usepackage{longtable}
\begin{document}

\begin{frontmatter}

\title{A partitioned fluid--structure interaction solver for two-phase sloshing and flexible spacecraft dynamics\tnoteref{arxiv-preprint}}
\tnotetext[arxiv-preprint]{Preprint version for arXiv.}

\author[isae,vki]{Umberto Zucchelli\corref{cor1}}
\ead{umberto.zucchelli2@isae-supaero.fr}
\author[vki]{Miguel Alfonso Mendez}
\author[isae]{Annafederica Urbano}
\author[esa]{Sebastien Vincent-Bonnieu}
\author[esa]{Piotr Wenderski}
\author[isae]{Francesco Sanfedino}

\address[isae]{F\'{e}d\'{e}ration ENAC ISAE-SUPAERO ONERA, Universit\'{e} de Toulouse,
  10 Av.\ \ Marc P\'{e}legrin, 31055 Toulouse, France}
\address[vki]{von Karman Institute for Fluid Dynamics, Environmental and Applied
  Fluid Dynamics Department, Waterloosesteenweg 72, B-1640 Sint-Genesius-Rode, Belgium}
\address[esa]{European Space Agency, ESTEC, Keplerlaan 1, 2201 AZ Noordwijk, the Netherlands}

\begin{abstract}
This paper presents a high-fidelity direct numerical simulation (DNS)–fluid–structure interaction (FSI) framework for rigid–liquid–flexible spacecraft dynamics under microgravity conditions. The liquid--gas flow is simulated with the incompressible two-phase solver implemented in DIVA, validated against FLUIDICS experiments conducted aboard the International Space Station (ISS). The flexible appendages are described by a rotating assumed-mode plate model that accounts for geometric stiffening. The fluid and structural operators are coupled through a Dirichlet–Neumann fixed-point algorithm with Aitken relaxation, and a closed-system mechanical energy balance is used as an a posteriori diagnostic to assess the energy imbalance of the partitioned discretisation. The coupling strategy is validated against an experimental free-decay sloshing benchmark, and its numerical consistency is assessed through spatial sensitivity studies of the energy-balance defect. Prescribed-motion, rigid open-loop, and flexible open-loop simulations of a spin-up manoeuvre are compared to isolate the effect of structural feedback on the sloshing response. Reduced liquid models identified from the different simulation architectures exhibit different predictive capabilities when embedded in the same rigid–flexible plant. A controller synthesized from the reduced model identified from the flexible simulation is replayed in the nonlinear CFD–FSI environment. The reduced model reproduces the principal attitude and actuator responses for the considered manoeuvre but does not recover the detailed nonlinear sloshing-load history. The framework provides a high-fidelity environment for analysing coupled spacecraft dynamics, identifying control-oriented models, and assessing reduced-model-based control strategies beyond their linear design representation.
\end{abstract}


\end{frontmatter}

\section{Introduction}
\label{sec:intro}

Next-generation spacecraft combine long-duration operations, repeated attitude manoeuvres, and substantial propellant masses. In such configurations, the liquid is not a passive payload: it redistributes inertia, generates time-dependent forces and torques, and interacts directly with the attitude dynamics. These effects are particularly pronounced in microgravity, where capillary forces strongly influence the interface shape and liquid redistribution, and where modest tank accelerations may produce large liquid reorientations and persistent sloshing loads \cite{Dalmon2019,simonini_2024}.

The difficulty increases when flexible appendages are present. Large solar panels, antennas, and lightweight deployable structures introduce low-frequency modes that may overlap with the dominant sloshing frequencies. Rotating flexible structures also exhibit gyroscopic, Coriolis, and geometric-stiffening effects, which modify their apparent stiffness and modal response during attitude manoeuvres \cite{Hoskoti2023,Rodrigues2024}. Recent studies on rigid--liquid--flexible spacecraft have shown that the coupled motion of propellant and appendages can modify the global dynamics and stability of the vehicle \cite{YuYue2023}. The problem is therefore intrinsically system-level. The liquid, the flexible structure, and the rigid body exchange loads on comparable time scales. As a consequence, a modelling assumption introduced in one subsystem can propagate through the coupled dynamics and affect the predicted spacecraft response.

A numerical framework for spacecraft sloshing must therefore account for the liquid motion and for its two-way interaction with the rigid and flexible spacecraft dynamics. The objective of this work is to develop such a framework for coupled simulations in microgravity. This objective can be approached through different modelling strategies, ranging from fast reduced-order representations used for system-level studies and control-oriented applications to higher-fidelity simulations designed to resolve the liquid interface dynamics and the associated wall loads \cite{ibrahim2005liquid, Dodge2000}. The appropriate modelling choice therefore depends on the required level of physical detail, and the acceptable computational cost.

For guidance, navigation, and control applications, the most common sloshing models remain frozen-mass, added-mass, and equivalent mechanical representations based on spring-mass and pendulum systems. These descriptions are attractive because they are inexpensive and can be embedded naturally in multibody and control formulations \cite{Abramson1966,Dodge2000,Gasbarri2016, capolupo2025equivalent}. Schott\'e and Ohayon \cite{SchotteOhayon2009} compared several modelling approaches for internal liquids in structural vibration analysis, ranging from frozen-liquid models to hydroelastic formulations with gravity. Their study illustrates why reduced models remain useful for system-level studies, but also clarifies their domains of validity. Frozen-mass models cannot describe relative liquid motion. Classical added-mass formulations do not resolve the free-surface dynamics. Equivalent mechanical models can reproduce selected sloshing modes, but their parameters are usually tied to a prescribed regime. Their predictive capability is therefore limited when the liquid undergoes large-amplitude motion, strong interface deformation, or large-scale capillary-driven redistribution within the tank under microgravity conditions. This limitation is important when the goal is not only to reproduce a dominant sloshing frequency, but also to predict the loads exchanged with a flexible spacecraft.

Continuum liquid models reduce part of this gap. Cho and Lee \cite{ChoLee2004} used a nonlinear finite-element formulation to study large-amplitude sloshing in baffled tanks. Yu and Yue \cite{YuYue2023} used isogeometric analysis with a level-set description to investigate a rigid--liquid--flexible spacecraft. These studies are important because they move beyond isolated sloshing estimates and address the coupled response of liquid-filled systems. In the present context, however, the fluid modelling assumptions remain central. Cho and Lee's potential-flow formulation assumes inviscid, irrotational motion, while the viscous level-set model of Yu and Yue neglects surface tension. Accurate modelling of microgravity sloshing requires the fluid model to capture capillary forces, contact-angle effects, interface deformation, and the resulting wall loads.

Particle methods provide another route for large-deformation free-surface flows. In the Particle Finite Element Method (PFEM), the fluid is represented by material nodes and the computational mesh is reconstructed as the domain evolves. This allows the free surface to be tracked by the moving discretization. Idelsohn et al. \cite{IdelsohnOnateDelPinCalvo2006} demonstrated the applicability of PFEM to free-surface flows and fluid--structure interaction involving large interface motion. Pan et al. \cite{PanCaoLi2020} later combined PFEM with multibody algorithms based on the absolute nodal coordinate formulation for rigid--liquid systems with free surfaces. Smoothed particle hydrodynamics has also become a widely used method for violent free-surface flows because it can handle large deformation, fragmentation, and impact. Recent developments have improved surface-tension and contact-angle treatments \cite{VergnaudOgerLeTouzeDeLeffeChiron2022}, particle regularization \cite{SunColagrossiMarroneAntuonoZhang2019,GaoFu2023}, and particle-based fluid--structure interaction \cite{MonteleoneBorinoNapoliBurriesci2022}. Nevertheless, long-time conservation remains a critical issue in violent free-surface SPH simulations \cite{PillotonSunZhangColagrossi2024}.
Together, these studies show that high-fidelity sloshing simulations require careful treatment of conservation, interface physics, and wall-load extraction.

Eulerian two-phase Navier--Stokes formulations offer a complementary reference framework. In contrast to particle and moving-mesh approaches such as SPH and PFEM, they resolve the governing equations on a fixed grid. Sharp-interface treatments preserve the pressure discontinuity at the liquid--gas interface, allowing capillary forces and the resulting wall loads to be accurately resolved \cite{Tryggvason2011}. Their main limitation is computational cost, which restricts their use in iterative design. Two-phase DNS combining sharp-interface methods with accurate capillary and wetting treatments can provide reliable, high-fidelity reference data for model validation, reduced-order identification, and control-oriented assessment. They can identify response frequencies and damping levels, test the validity range of equivalent mechanical models \cite{Abramson1966,Dodge2000}, and evaluate the loads exchanged with the spacecraft when the coupled dynamics are restored. Direct numerical simulation is therefore used here as a reference validation environment, not as a real-time control model.

On this basis, the present work develops a high-fidelity DNS--FSI framework for rigid--liquid--flexible spacecraft dynamics in microgravity. The configuration consists of a rigid spacecraft hub, a partially filled spherical tank, and flexible panels. The liquid--gas flow is simulated using the incompressible two-phase solver implemented in the \textsc{DIVA} code. The solver represents the liquid--gas interface with a level-set method and enforces sharp pressure jumps through ghost-fluid treatments. Dalmon et al. \cite{Dalmon2019} compared \textsc{DIVA} simulations with \textsc{FLUIDICS} measurements obtained on board the ISS for spherical tanks under prescribed rotational manoeuvres. The comparison included the liquid configuration and the time-dependent force and torque exerted on the tank wall, and showed that the DNS reproduced the dominant oscillation frequencies and the measured load trends in the investigated Bond-number range. The present work builds on this validation and extends the use of \textsc{DIVA} from prescribed tank motion to two-way coupled spacecraft dynamics, where the motion of the spacecraft modifies the liquid forcing and the resulting sloshing loads feed back into the rigid--flexible response.

The fluid solver is coupled with a rigid-body attitude model and a reduced structural model for the flexible appendages. The flexible panels are represented through a modal formulation that retains the dominant structural dynamics while accounting for rotation-induced geometric-stiffening effects \cite{VallesSanchez2025}. The fluid and rigid--flexible solvers exchange tank kinematics and hydrodynamic loads through a partitioned coupling strategy. This formulation is first assessed against available experimental data from the literature, so that the coupled exchange of motion and loads can be verified before addressing the spacecraft application. It is then applied to representative microgravity spin-up manoeuvres involving simultaneous sloshing, flexible-panel vibration, and rigid-body attitude motion.

The contribution of the paper is threefold. First, it formulates a high-fidelity DNS--FSI framework for spacecraft dynamics in microgravity, with two-way coupling between rigid-body motion, nonlinear two-phase sloshing, and flexible-panel dynamics. The formulation couples a \textsc{FLUIDICS}-validated two-phase DNS solver with a rotating flexible-panel model through a Dirichlet--Neumann fixed-point algorithm. Second, it validates the coupled exchange against available experimental data from the literature and introduces a closed-system mechanical-energy balance to assess the numerical consistency of the coupled simulations. Third, it applies this framework to representative spin-up manoeuvres and analyses the modal interaction between nonlinear sloshing, rotating flexible panels, and rigid-body attitude motion, with the broader objective of supporting reduced-order model development and control-validation studies.

The remainder of the paper is organised as follows.
\Cref{sec:methodology} presents the DNS--FSI computational framework: the two-phase fluid model and the numerical methods used for its solution, the flexible-panel model, the rigid-body attitude dynamics, and the Dirichlet–Neumann partitioned coupling algorithm. 
\Cref{sec:energy} defines the closed-system energy balance and the a posteriori diagnostic used for verification.
\Cref{sec:config} defines the reference spacecraft, manoeuvre, and numerical settings.
\Cref{sec:control} introduces the control-oriented liquid surrogate and LTI plant used for LQG synthesis.
\Cref{sec:results} first assesses the translational coupling against the Peterson experiment \cite{peterson1989nonlinear}, then reports the energy-balance sensitivity study, compares prescribed-motion, rigid-FSI, and flexible-FSI responses, and finally evaluates the LQG controller in closed-loop DNS--FSI replay.
\Cref{sec:conclusions} summarises the main findings and implications for reduced-order modelling and control validation.

\section{Partitioned fluid--structure coupling}
\label{sec:methodology}
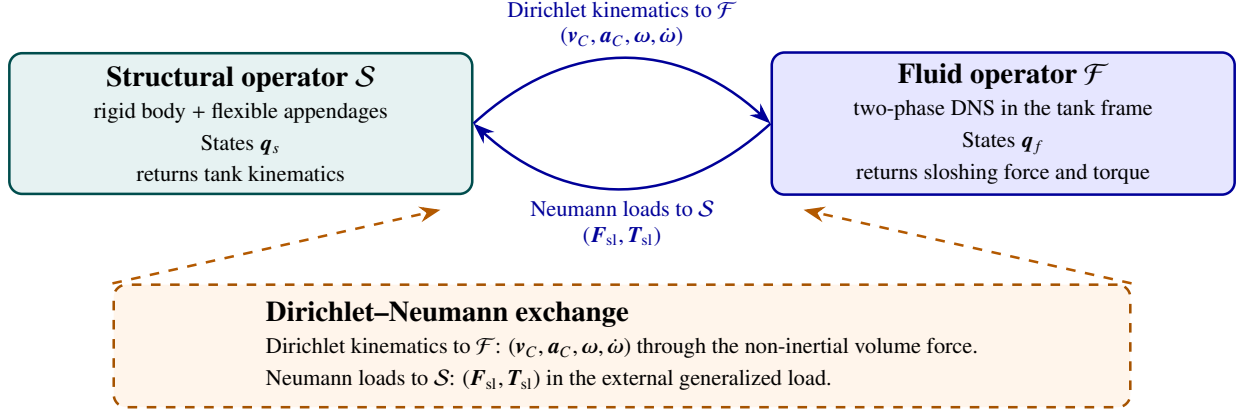
\begin{figure*}[!ht]
\centering
\resizebox{\textwidth}{!}{
\begin{tikzpicture}[
  block/.style={draw, rounded corners=4pt, thick, minimum width=4.7cm,
    minimum height=1.55cm, align=center, font=\small},
  structure/.style={block, fill=teal!10, draw=teal!60!black,
    minimum width=5.1cm},
  fluid/.style={block, fill=blue!10, draw=blue!60!black,
    minimum width=5.1cm},
  policy/.style={draw, dashed, rounded corners=4pt, thick,
    fill=orange!8, draw=orange!65!black, minimum width=11.2cm,
    minimum height=1.35cm, align=left, font=\small},
  note/.style={font=\scriptsize, fill=white, inner sep=1.5pt, align=center},
  >=Stealth]

  \node[structure] (struct) at (-4.2,0.9)
    {\textbf{Structural operator $\mathcal{S}$}\\[-1pt]
     \scriptsize rigid body + flexible appendages \\[-1pt]
     \scriptsize States $\vect{q}_s$ \\[-1pt]
     \scriptsize returns tank kinematics};
  \node[fluid] (diva) at (4.2,0.9)
    {\textbf{Fluid operator $\mathcal{F}$}\\[-1pt]
     \scriptsize two-phase DNS in the tank frame\\[-1pt]
     \scriptsize States $\vect{q}_f$\\[-1pt]
     \scriptsize returns sloshing force and torque};
  \node[policy] (cpl) at (0,-1.55)
    {\textbf{Dirichlet--Neumann exchange}\\[-1pt]
     \scriptsize Dirichlet kinematics to $\mathcal{F}$: $(\vect{v}_C, \vect{a}_C, \vect{\omega},\dot{\vect{\omega}})$ through the non-inertial volume force.\\[-1pt]
     \scriptsize Neumann loads to $\mathcal{S}$: $(\vect{F}_\mathrm{sl},\vect{T}_\mathrm{sl})$ in the external generalized load. };

  \draw[->, thick, blue!60!black]
    (struct.east) .. controls (-0.6,1.85) and (0.6,1.85) .. (diva.west)
    node[note, midway, above=2pt]
      {Dirichlet kinematics to $\mathcal{F}$\\$(\vect{v}_C, \vect{a}_C, \vect{\omega},\dot{\vect{\omega}})$};
  \draw[->, thick, blue!60!black]
    (diva.west) .. controls (0.6,-0.05) and (-0.6,-0.05) .. (struct.east)
    node[note, midway, below=2pt]
      {Neumann loads to $\mathcal{S}$\\$(\vect{F}_\mathrm{sl},\vect{T}_\mathrm{sl})$};

  \draw[->, thick, dashed, orange!70!black] (cpl.north west) -- (-2.0,-0.05);
  \draw[->, thick, dashed, orange!70!black] (cpl.north east) -- (2.0,-0.05);
\end{tikzpicture}
}
\caption{Operator-level workflow of the partitioned DNS--FSI framework. The structural operator supplies the tank kinematics imposed on the fluid problem; the fluid operator returns the wall loads applied to the structural problem.}
\label{fig:framework}
\end{figure*}

The coupled problem considered in this work combines three physical subsystems: an incompressible immiscible two-phase flow in a partially filled spherical tank, the rigid-body attitude dynamics of the spacecraft, and the flexible dynamics of the solar panels. The liquid is modeled in the tank-attached frame, where the prescribed or computed angular kinematics generate the non-inertial forcing seen by the fluid; in return, the fluid solver provides the sloshing loads that act on the structural subsystem. This modelling is written in terms of a fluid operator \(\mathcal{F}\) and a rigid--flexible structural operator \(\mathcal{S}\). This operator view is summarised in \Cref{fig:framework}.

\begin{figure}[!ht]
\centering
\includegraphics[width=\linewidth]{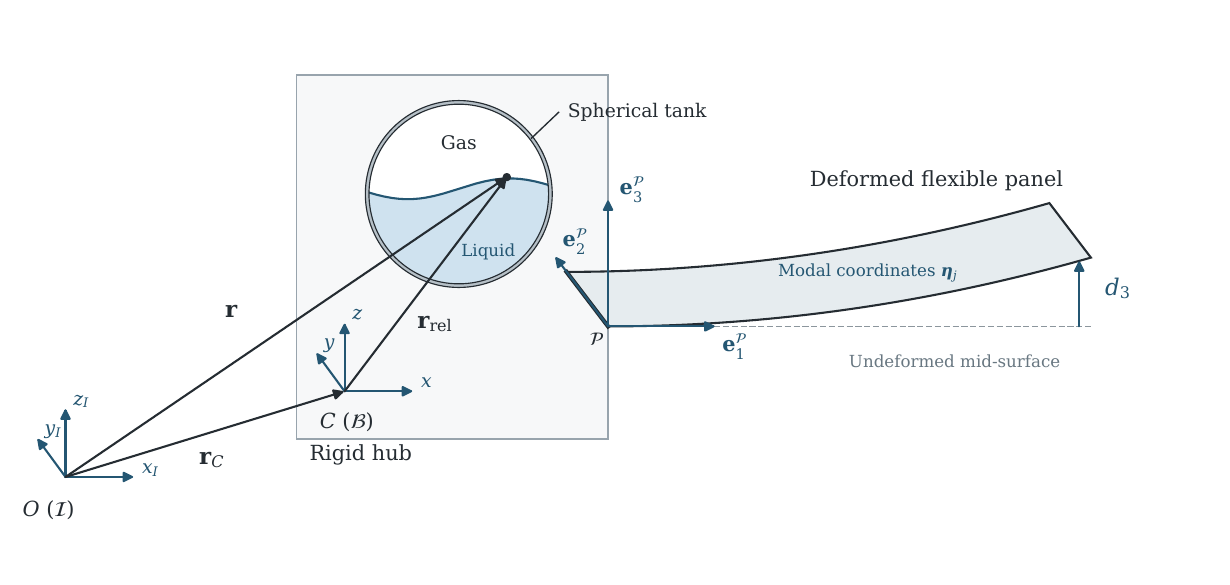}
\caption{Reference frames and position vectors on a representative spacecraft schematic. $\mathcal I$ is the inertial frame, $\mathcal B$ is the body frame attached to the hub at $C$, and $\mathcal P$ is the panel frame attached to the panel root.  Vectors $\vect r_C$ and $\vect r$ locate $C$ and a fluid point from the inertial origin; the arrow from $C$ to that point is $\vect r_{\mathrm{rel}}=\vect r-\vect r_C$. The panel basis vectors $\vect e_1^{\mathcal P}$, $\vect e_2^{\mathcal P}$ and $\vect e_3^{\mathcal P}$ point along the undeformed length, width and normal, respectively, whereas $\vect\eta_j$ contains the modal coordinates of panel $j$.}
\label{fig:anna_frames}
\end{figure}
\subsection{Dirichlet--Neumann partitioned coupling}
\label{sec:coupling}
\noindent The DNS--FSI problem is advanced by fixed-point iterations within each physical time step. The structure supplies the tank kinematics to the fluid (Dirichlet data), while the fluid returns the hydrodynamic force and torque as dynamic loads on the structure (Neumann data) \cite{Farhat1998,Bungartz2016}. 

Let
\begin{equation}
  \vect{q}_f = (\vect{u},\phi,p),
  \qquad
  \vect{q}_s =
	(\vect{q}_{BI}, \vect{\omega}, \vect{v}_C, 
    \vect{\eta}_1,\dot{\vect{\eta}}_1,
    \vect{\eta}_2,\dot{\vect{\eta}}_2),
    \qquad
	\vect{\ell}_\mathrm{sl}=
	\begin{bmatrix}\vect{F}_\mathrm{sl}^T&\vect{T}_\mathrm{sl}^T\end{bmatrix}^T
\end{equation}
collect the fluid state $\vect{q}_f$, the structural state $\vect{q}_s$, and the hydrodynamic sloshing wrench $\vect{\ell}_\mathrm{sl}$, respectively. The structural state contains the body-to-inertial unit quaternion \(\vect{q}_{BI}\), the body angular velocity \(\vect{\omega}\), the translational velocity \(\vect{v}_C\), and the modal coordinates and velocities \((\vect{\eta}_j,\dot{\vect{\eta}}_j)\) of panel \(j\). The hub-fixed reference point \(C\) coincides with the nominal centre of gravity in the spacecraft configuration; its inertial acceleration expressed in the body frame, \(\vect{a}_C=\dot{\vect{v}}_C+\vect\omega\times\vect v_C\), is obtained by solving the structural equations of motion. The force \(\vect{F}_\mathrm{sl}\) and torque \(\vect{T}_\mathrm{sl}\) act on the tank, are expressed in the body frame, and the torque is taken about \(C\), as illustrated in \Cref{fig:anna_frames}. Over one physical time step $t^n\rightarrow t^{n+1}$, the two solvers are written as
\begin{align}
  \mathcal{F}:\quad
  \vect{q}_f^{n+1}
  &=
  \mathcal{F}
  \left(
  \vect{q}_f^n;
  \vect{v}_C,\vect{a}_C,
  \vect{\omega},\dot{\vect{\omega}}
  \right),
  &
  \vect{\ell}_\mathrm{sl}^{n+1}
  &=
  \mathcal{L}
  \left(
  \vect{q}_f^{n+1}
  \right),
  \label{eq:fluid_op}
  \\[3pt]
  \mathcal{S}:\quad
  \vect{q}_s^{n+1}
  &=
  \mathcal{S}
  \left(
  \vect{q}_s^n;
  \vect{\ell}_\mathrm{sl}
  \right),
  &
  \left(
  \vect{v}_C^{n+1},
  \vect{a}_C^{n+1},
  \vect{\omega}^{n+1},
  \dot{\vect{\omega}}^{n+1}
  \right)
  &=
  \mathcal{K}
  \left(
  \vect{q}_s^{n+1}
  \right),
  \label{eq:struct_op}
\end{align}
where $\mathcal{L}$ denotes the wall-load extraction operator and $\mathcal{K}$ extracts the translational and angular kinematics required by the fluid solver.  The flexible coordinates remain internal to $\mathcal{S}$ and affect the fluid through the resulting tank kinematics. The general exchange comprises the translational and angular velocities and accelerations, together with the six-component hydrodynamic wrench. The spacecraft calculations in \Cref{sec:config} impose \(\vect{a}_C=\vect{0}\) as a constraint on the linear dynamics and retain the rotational pair \((\dot{\vect{\omega}},\vect{T}_\mathrm{sl})\). The Peterson benchmark in \Cref{sec:validation} instead retains the translational pair \((a_{C,x},F_{\mathrm{sl},x})\).
A consistent partitioned solution at time $t^{n+1}$ satisfies the fixed-point problem
\begin{equation}
  \vect{\ell}_\mathrm{sl}^{\star}
  =
  \mathcal{H}
  \left(
  \vect{\ell}_\mathrm{sl}^{\star}
  \right),
  \qquad
  \mathcal{H}
  =
  \mathcal{L}
  \circ
  \mathcal{F}
  \circ
  \mathcal{K}
  \circ
  \mathcal{S}.
  \label{eq:fixed_point_map}
\end{equation}
Equivalently, for the relaxed load estimate $\vect{\ell}_\mathrm{sl}^{(k)}$ at coupling iteration \(k\),
	\begin{align}
	\vect{q}_s^{(k+1)}
	&=
	\mathcal{S}
	\left(
	\vect{q}_s^n;
	\vect{\ell}_\mathrm{sl}^{(k)}
	\right),
	\label{eq:fixed_point_struct}
	\\
	\left(
	\vect{v}_C^{(k+1)},
	\vect{a}_C^{(k+1)},
	\vect{\omega}^{(k+1)},
	\dot{\vect{\omega}}^{(k+1)}
	\right)
  &=
  \mathcal{K}
  \left(
  \vect{q}_s^{(k+1)}
  \right),
  \label{eq:fixed_point_kin}
  \\
  \vect{q}_f^{(k+1)}
  &=
  \mathcal{F}
  \left(
  \vect{q}_f^n;
  \vect{v}_C^{(k+1)},
  \vect{a}_C^{(k+1)},
  \vect{\omega}^{(k+1)},
  \dot{\vect{\omega}}^{(k+1)}
  \right),
  \label{eq:fixed_point_fluid}
  \\
  \widehat{\vect{\ell}}_\mathrm{sl}^{(k+1)}
  &=
  \mathcal{L}
  \left(
  \vect{q}_f^{(k+1)}
  \right).
  \label{eq:fixed_point_load}
\end{align}
A hat denotes the raw wrench returned by the most recent fluid solve, whereas the corresponding symbol without a hat denotes the relaxed wrench supplied to the next structural iterate; superscripts \((k)\) and \(n\) identify coupling iterations and accepted physical time levels, respectively. In the present work, \Cref{eq:fixed_point_map} is solved by a strong partitioned coupling in which the structural solve and the full DNS step are repeated until the exchanged load and the resulting structural acceleration are mutually consistent at $t^{n+1}$. Operationally, the algorithm is partitioned because the fluid and structural subproblems remain distinct, and implicit at the coupling level because the interface equilibrium is enforced by sub-iterations rather than by a single lagged exchange. \Cref{alg:strong_coupling} summarizes the complete procedure.

\noindent The coupling algorithm solves \Cref{eq:fixed_point_map} by sub-iterations within each physical time step. At every coupling iteration, the structural state is recomputed from the current load estimate, the tank kinematics are updated, and the complete Runge--Kutta DNS step is repeated from the beginning-of-step fluid state. Hence, the fluid state and the extracted load are consistent with the current tank kinematics. This checkpoint-and-resolve procedure is standard for strong partitioned FSI coupling \cite{Degroote2010,Bungartz2016}.
The fixed-point residual \(r_*\) is monitored using the exchanged-wrench residual \(r_\ell\) and the structural-acceleration residual \(r_a\). Here, \(\vect{a}_s\) collects the active structural-acceleration components supplied to the fluid solver; it reduces to \(a_{C,x}\) for the Peterson benchmark and to \(\dot{\vect{\omega}}\) for the spacecraft simulations. The two residuals are :
\begin{equation}
	r_\ell^{(k)} =
	\frac{ \left\| \widehat{\vect{\ell}}_\mathrm{sl}^{(k+1)} - \vect{\ell}_\mathrm{sl}^{(k)} \right\|
	}{
	\left\| \widehat{\vect{\ell}}_\mathrm{sl}^{(k+1)} \right\| + \varepsilon_0
},
\qquad
r_a^{(k)} =
\frac{ \left\| \vect{a}_s^{(k+1)} - \vect{a}_s^{(k)} \right\|
}{
\left\| \vect{a}_s^{(k+1)} \right\| + \varepsilon_0
},
\label{eq:residuals}
\end{equation}
with $\varepsilon_0=10^{-12}$. Convergence is declared when
\begin{equation}
  \max
  \left(
  r_\ell^{(k)},r_a^{(k)}
  \right)
  <
  \varepsilon_\mathrm{FSI},
  \qquad
  \varepsilon_\mathrm{FSI}=10^{-6}.
  \label{eq:fsi_convergence}
\end{equation}

\noindent Aitken dynamic relaxation \cite{Kuettler2008,Degroote2010} is applied to the exchanged wrench. It damps oscillatory fixed-point corrections and accelerates convergence when successive corrections become aligned. Defining the raw load increment
\begin{equation}
  \Delta\vect{\ell}^{(k)}
  =
  \widehat{\vect{\ell}}_\mathrm{sl}^{(k+1)} - \vect{\ell}_\mathrm{sl}^{(k)},
\label{eq:load_increment}
\end{equation}
the load applied to structural iteration \(k+1\) is
\begin{equation}
  \vect{\ell}_\mathrm{sl}^{(k+1)}
  =
  \vect{\ell}_\mathrm{sl}^{(k)} + \theta^{(k)}\Delta\vect{\ell}^{(k)} .
  \label{eq:aitken}
\end{equation}
The first correction uses \(\theta^{(0)}=\theta_0=0.7\). For $k\geq 1$, the dimensionless relaxation parameter is updated as
\begin{equation}
  \theta^{(k)}
  =
  -\theta^{(k-1)} \frac{ \left\langle \Delta\vect{\ell}^{(k-1)},
	\Delta\vect{\ell}^{(k)} - \Delta\vect{\ell}^{(k-1)} \right\rangle
  }{
  \left\| \Delta\vect{\ell}^{(k)} - \Delta\vect{\ell}^{(k-1)} \right\|^2
  },
  \qquad
  \theta^{(k)}
  \leftarrow
  \max\left(0.1,\min(1,\theta^{(k)})\right).
  \label{eq:aitken_update}
\end{equation}
The clipping interval \([0.1,1]\) prevents over-relaxation while preserving a non-zero correction when successive increments become nearly collinear. The complete step-by-step sequence over one physical time step is summarized in \Cref{alg:strong_coupling}.

\begin{algorithm}[!ht]
\caption{Dirichlet-Neumann fixed-point coupling over one physical time step.}
\label{alg:strong_coupling}
\begin{algorithmic}[1]
\Require $\vect{q}_f^n,\vect{q}_s^n,\vect{\ell}_\mathrm{sl}^{n}$
\State Save the beginning-of-step fluid and structural states.
\State Set $k=0$, $\vect{\ell}_\mathrm{sl}^{(0)}=\vect{\ell}_\mathrm{sl}^{n}$,
and $\theta^{(0)}=\theta_0$.
\Repeat
  \State Restore \(\vect{q}_s^n\) and solve structural problem $\vect{q}_s^{(k+1)}
    =
    \mathcal{S} \left(\vect{q}_s^n;\vect{\ell}_\mathrm{sl}^{(k)}\right).$
	\State Extract $\left(\vect{v}_C^{(k+1)},\vect{a}_C^{(k+1)},\vect{\omega}^{(k+1)},\dot{\vect{\omega}}^{(k+1)}\right)
    =\mathcal{K}\left(\vect{q}_s^{(k+1)}\right).$
  \State Restore \(\vect{q}_f^n\) and solve fluid $\vect{q}_f^{(k+1)}
    =
    \mathcal{F} \left(\vect{q}_f^n;\vect{v}_C^{(k+1)},\vect{a}_C^{(k+1)},\vect{\omega}^{(k+1)},\dot{\vect{\omega}}^{(k+1)}\right).$
	\State Extract $\widehat{\vect{\ell}}_\mathrm{sl}^{(k+1)}
	=\mathcal{L}\left(\vect{q}_f^{(k+1)}\right).$
	\State Evaluate $r_\ell^{(k)}$ and $r_a^{(k)}$ from
  \Cref{eq:residuals}.
  \If{$\max(r_\ell^{(k)},r_a^{(k)})<\varepsilon_\mathrm{FSI}$}
    \State Accept
    $\vect{q}_f^{n+1}=\vect{q}_f^{(k+1)}$,
    $\vect{q}_s^{n+1}=\vect{q}_s^{(k+1)}$, and
    $\vect{\ell}_\mathrm{sl}^{n+1}=\widehat{\vect{\ell}}_\mathrm{sl}^{(k+1)}$.
    \State \textbf{break}
  \EndIf
  \State Compute $\Delta\vect{\ell}^{(k)}$ from
  \Cref{eq:load_increment}.
  \If{$k\geq 1$}
    \State Update $\theta^{(k)}$ from \Cref{eq:aitken_update}.
  \EndIf
  \State Update the relaxed load estimate using \Cref{eq:aitken}.
  \State $k\gets k+1$.
\Until{$k=k_\mathrm{max}$}
\If{$\max(r_\ell^{(k)},r_a^{(k)})\geq\varepsilon_\mathrm{FSI}$}
  \State Retain the final computed iterates as
  \(\vect{q}_f^{n+1}\), \(\vect{q}_s^{n+1}\), and
  \(\vect{\ell}_\mathrm{sl}^{n+1}\), and flag the step as non-converged.
\EndIf
\State \Return
$\vect{q}_f^{n+1},\vect{q}_s^{n+1},\vect{\ell}_\mathrm{sl}^{n+1}$.
\end{algorithmic}
\end{algorithm}

\noindent The coupling algorithm therefore enforces consistency between the hydrodynamic wrench, the generalized translational--rotational--modal acceleration, and the non-inertial forcing used by the DNS solver within each physical time step.

\subsection{Fluid subsystem $\mathcal{F}$}
\label{sec:diva}
	
The fluid operator \(\mathcal{F}\) used for the sloshing simulations is the incompressible two-phase solver implemented in the in-house \textsc{DIVA} code. The solver has been validated for isothermal and phase-change two-phase flows \mbox{\cite{Tanguy2005, Lalanne2015, Lepilliez2016, RuedaVillegas2017, Urbano2018}} and, against experimental data for isothermal sloshing in microgravity from the \textsc{FLUIDICS} experiment aboard the International Space Station \mbox{\cite{Mignot2017, Dalmon2019, DalmonThesis2019}}. The fluid is modelled as an incompressible, immiscible, isothermal two-phase flow, in a three-dimensional Cartesian grid attached to the tank reference frame.An immersed-boundary method (IBM) is used to describe complex geometries on this grid \cite{Lepilliez2016}. 

A one-fluid formulation is adopted for the liquid and gas phases. The flow is described by the fluid velocity relative to the tank \(\vect{u}\), the pressure \(p\), and a signed-distance level-set field \(\phi\), defined such that \(\phi>0\) in the liquid and \(\phi<0\) in the gas. All field variables implicitly depend on the body-frame coordinates \(\vect r_{\rm rel}\), measured from \(C\), and time \(t\). The tank is fixed in this frame. In terms of the inertial position vectors in \Cref{fig:anna_frames}, \(\vect r_{\rm rel}=\mat R_{BI}^{T}(\vect r^{\mathcal I}-\vect r_C^{\mathcal I})\), where \(\mat R_{BI}\) maps body-frame components to the inertial frame and corresponds to the quaternion \(\vect q_{BI}\). The density \(\rho(\phi)\) and dynamic viscosity \(\mu(\phi)\) take their respective bulk values \((\rho_\ell,\mu_\ell)\) in the liquid and \((\rho_g,\mu_g)\) in the gas. The governing equations read
\begin{align}
\divop \vect{u} &= 0,
\label{eq:div_free_fluid}\\
\rho(\phi)\left(
\frac{\partial \vect{u}}{\partial t} + \vect{u}\cdot\grad \vect{u}
\right)
&= -\grad p
+ \divop\!\left(2\mu(\phi)\mat{D}(\vect{u})\right)
+ \rho(\phi)\vect{f}_\mathrm{vol},
\label{eq:one_fluid_ns}\\
\frac{\partial \phi}{\partial t}
+ \vect{u}\cdot\grad \phi &=0,
\label{eq:level_set_continuous}
\end{align}
where \(\mat{D}(\vect{u})=(\grad\vect{u}+\grad\vect{u}^{T})/2\) is the rate-of-strain tensor, and \(\vect{f}_\mathrm{vol}\) is the non-inertial volume acceleration defined below. Surface tension is incorporated as a sharp capillary jump in the pressure at the liquid--gas interface $\Gamma(t)=\{\vect{r}:\phi(\vect{r},t)=0\}$ by the Ghost Fluid Method \cite{Fedkiw1999,Liu2000}.
\begin{equation}
[p]_{\ell g}=p_\ell-p_g=-\sigma\kappa,
\qquad
\kappa=\divop\left(\frac{\grad\phi}{|\grad\phi|}\right),
\label{eq:pressure_jump_convention}
\end{equation}
Here \([p]_{\ell g}=p_\ell-p_g\) is the liquid-to-gas pressure jump, \(\sigma\) is the surface-tension, \(\kappa\) is the interface curvature, and \(\vect n=\grad\phi/|\grad\phi|\) is the unit normal directed from gas to liquid.
Pressure and level-set variables are stored at cell centres of a staggered MAC grid and velocity components at face centres \cite{HarlowWelch1965}. Convective terms are discretised with fifth-order WENO-Z advection \cite{Jiang1996,BorgesEtAl2008,Shu2009WENO}; the level-set field is redistanced after each physical step using the Hamilton--Jacobi procedure of Sussman et al. \cite{Sussman1994}.

 \noindent The tank wall is represented by a second signed-distance field \(\phi_s\), positive in the fluid domain, and a cut-cell immersed-boundary method following the Cartesian-grid formulations of \cite{Ye1999, TsengFerziger2003, MittalIaccarino2005, TairaColonius2007,Lepilliez2016}. The resulting open-face fractions and cut-cell volumes are used consistently in the convective fluxes, pressure operator, velocity correction, and load reconstruction; the pressure stencil near cut cells follows the subcell treatment of \cite{Gibou2002,Ng2009}. The immersed-boundary velocity correction enforces no slip at the reconstructed wall, and stencils crossing the solid ghost band are completed with the PDE-based extension of \citet{Aslam2004}. At the triple contact line, the prescribed contact angle \(\theta_c\) is imposed through
\begin{equation}
	\vect{n}_s\cdot\grad\phi=\cos\theta_c,
	\qquad \vect{n}_s=\frac{\grad\phi_s}{|\grad\phi_s|}.
\label{eq:contact_angle}
\end{equation}
where \(\vect{n}_s\) is the unit normal directed from the solid into the fluid. In the spacecraft simulations, a null contact angle $\theta_c=0$ is imposed at the triple line, i.e. a perfectly wetting condition in which a thin liquid film may remain on the tank wall. This wetting condition follows the FLUIDICS configuration \cite{Dalmon2019}.

\noindent The interface and momentum equations are advanced with a second-order Runge--Kutta scheme \cite{Shu1988}. Each stage transports \(\phi\), updates the contact-angle extension, forms a viscous--convective predictor, solves the variable-coefficient pressure equation, and corrects the velocity with a non-incremental Chorin--Temam projection \cite{Chorin1967,Guermond2006,Sussman2007}. Viscous diffusion is treated semi-implicitly with the conservative GFCMI operator \cite{Lepilliez2016}. The pressure equation includes the ghost-fluid capillary jump and is solved with black-box multigrid \cite{Dendy1982,MacLachlan2008}. 

\noindent The tank kinematics supplied by \(\mathcal S\) constitute the Dirichlet data in \Cref{fig:framework}. In the tank-attached frame, they enter \(\mathcal F\) through the non-inertial volume acceleration, 
\begin{equation}
\vect{f}_\mathrm{vol}
=\vect{g}-\vect{a}_C-2\vect{\omega}\times\vect{u}
-\dot{\vect{\omega}}\times\vect{r}_\mathrm{rel}
-\vect{\omega}\times
 (\vect{\omega}\times\vect{r}_\mathrm{rel}),
\label{eq:non_inertial_force}
\end{equation}
Here \(\vect g=\mat R_{BI}^{T}\vect g^{\mathcal I}\) is the gravitational acceleration expressed in the body frame, and \(\vect r_{\rm rel}\) is the body-frame position relative to \(C\), as defined above. The four terms after gravity are, respectively, the translational, Coriolis, Euler, and centrifugal accelerations. The spacecraft cases use \(\vect{g}=\vect{0}\). Thus, a change in structural acceleration modifies the pressure, interface, and wall loads within the same physical step. In \textsc{DIVA}, the adaptive physical time step \(\Delta t\) combines the convective and capillary stability limits through their reciprocal sum, with \(\mathrm{CFL}_\mathrm{conv}=0.5\) and \(\mathrm{CFL}_\sigma=0.25\), respectively. These coefficients and the remaining numerical settings are collected in \Cref{tab:params}.

The fluid solver \(\mathcal F\) returns to the structural solver \(\mathcal S\) the resultant hydrodynamic force and torque acting on the tank:
\begin{align}
\vect F_{\rm sl} &=\int_{\Gamma_w}\left(p\vect n_w-\mat\tau\vect n_w\right)\,dA,\\
\vect T_{\rm sl} &=\int_{\Gamma_w}\vect r_{\rm rel}\times \left(p\vect n_w-\mat\tau\vect n_w\right)\,dA,
\label{eq:anna_wall_resultants}
\end{align}
where \(\Gamma_w\) is the tank wall, \(\vect n_w=-\vect n_s\) points from the fluid into the solid, and \(\mat\tau=2\mu\mat D(\vect u)\) is the viscous stress tensor. With this convention, \(p\vect n_w\) and \(-\mat\tau\vect n_w\) are the pressure and viscous effects exerted by the fluid on the tank.

These surface integrals define the physical loads exchanged with the structure. In DIVA, the tank wall is represented on the Cartesian grid by the immersed-boundary method introduced above. The corresponding force and torque are evaluated from the fluid momentum changes associated with the pressure projection and the enforcement of the wall velocity, together with the viscous wall-stress contribution. The resulting force and torque are expressed in the body frame, with torque taken about \(C\), and form the wrench \(\vect\ell_{\rm sl}=[\vect F_{\rm sl}^{T},\vect T_{\rm sl}^{T}]^{T}=\mathcal L(\vect q_f)\) returned to \(\mathcal S\) in \Cref{eq:fluid_op}.

\subsection{Structural subsystem $\mathcal{S}$}
\label{sec:flex_model}

The structural operator \(\mathcal{S}\) combines the rigid-body attitude dynamics of the hub with two flexible panels. Their dynamics is represented by the reduced-order plate model presented in \cite{VallesSanchez2025}, previously validated against finite-element results. It retains the dominant bending modes and the frequency shifts introduced by gyroscopic terms and centrifugal stiffening.\\
Each panel is modelled as a clamped-free Kirchhoff-Love plate of length $L$, width $D$, thickness $h$, and surface density $\rho_\delta=\rho_p h$. 
The panel deformation is described in the root-attached frame $\mathcal P$ shown in \Cref{fig:anna_frames}, whose basis vectors $\vect e_1^{\mathcal P}$, $\vect e_2^{\mathcal P}$ and $\vect e_3^{\mathcal P}$ point along the panel length, width and normal, respectively, with coordinates $\xi \in [0,L]$ along the length and $\zeta \in [0,D]$ along the width. The undeformed mid-surface spans the domain $A = [0,L] \times [0,D]$, with area element $dA = d\xi\,d\zeta$. Under thin-plate theory, in-plane membrane deformations are neglected, so that the elastic displacement reduces to the out-of-plane transverse deflection $d_3(\xi,\zeta,t)$. The deflection is approximated by a modal expansion
\begin{equation}
  d_3(\xi,\zeta,t)
  =
  \sum_{k=1}^{N_m}\eta_k(t)\,W_k(\xi,\zeta),
  \qquad
    W_k(\xi,\zeta) =  \Phi_{m(k)}(\xi)\,\Psi_{n(k)}(\zeta),
  \label{eq:modal_expansion}
\end{equation}
where $\vect{\eta}=[\eta_1,\ldots,\eta_{N_m}]^T$ is the vector of modal coordinates, and the trial functions $W_k$ are formed by products of clamped-free beam modes $\Phi_m(\xi)$ along the length and free-free beam modes $\Psi_n(\zeta)$ across the width. In this work, $N_\Phi=2$ and $N_\Psi=2$, yielding $N_m=N_\Phi N_\Psi=4$ modes per panel.
The modal equations of motion in the non-inertial frame $\mathcal{P}$ are
\begin{equation}
  \mat{M}\ddot{\vect{\eta}}
  +
  \mat{D}\dot{\vect{\eta}}
  +
  \left[
  \mat{K}_\mathrm{sp}
  +
  \mat{K}_d(\vect{\omega}^{\mathcal{P}},
  \dot{\vect{\omega}}^{\mathcal{P}})
  \right]\vect{\eta}
  =
  -\mat{L}^T\vect{a}^{\mathcal{P}}
  -\mat{P}^T\dot{\vect{\omega}}^{\mathcal{P}},
  \label{eq:modal_eom}
\end{equation}
where $\mat{M}$ is the modal mass matrix, $\mat{K}_\mathrm{sp}$ is the linear plate stiffness matrix, $\mat{K}_d$ is the spin-dependent geometric stiffness, and \(\mat{L}\) and \(\mat{P}\) couple the panel modes to the acceleration of \(C\) expressed in the panel frame, \(\vect a^{\mathcal P}=\mat R_{PB}^{T}\vect a_C\), and angular acceleration \(\dot{\vect{\omega}}^{\mathcal P}\), respectively. The dynamic stiffness depends on the angular velocity $\vect{\omega}^{\mathcal{P}}$ and angular acceleration $\dot{\vect{\omega}}^{\mathcal{P}}$ expressed in the panel frame and accounts for the centrifugal and Euler contributions induced by the foreshortening kinematics. The damping matrix combines gyroscopic and structural dissipation,
\begin{equation}
  \mat{D}
  =
  2\mat{D}_\mathrm{gyro}(\vect{\omega}^{\mathcal{P}}) + \mat{D}_\mathrm{Ray},
  \qquad
  \mat{D}_\mathrm{Ray}
  =
  \alpha_R\mat{M} + \beta_R \left[\mat{K}_\mathrm{sp} + \mat{K}_d \right],
  \label{eq:panel_damping}
\end{equation}
with $\mat{D}_\mathrm{gyro}$ the gyroscopic operator and $\alpha_R,\beta_R$ the Rayleigh damping coefficients. \\

Let $\vect{r}_{P0}^{\mathcal{P}}=[A_l,B_l,C_l]^T$ denote the offset vector from the hub reference point to the panel root, expressed in the local panel frame $\mathcal{P}$. For a panel point whose transverse displacement is along the local unit vector \(\vect{e}^{\mathcal P}_3\), the translational-acceleration coupling matrix \(\mat{L}\in\mathbb{R}^{3\times N_m}\) and angular-acceleration coupling matrix $\mat{P}\in\mathbb{R}^{3\times N_m}$ are
\begin{equation}
L_{ij}=\int_A \rho_\delta\,(\vect{e}^{\mathcal P}_i\!\cdot\vect{e}^{\mathcal P}_3)\,W_j\,dA,
\qquad i=1,2,3,
\label{eq:L_matrix}
\end{equation}
and
\begin{align}
  P_{1j}
  &=
  \int_A \rho_\delta \left(B_l+\zeta-\frac{D}{2}\right) W_j\,dA, \notag\\
  P_{2j}
  &=
  -\int_A \rho_\delta \left(A_l+\xi\right) W_j\,dA, \qquad
  P_{3j}=0 .
  \label{eq:P_matrix}
\end{align}
The root offset is included in $\mat P$, so $\vect a^{\mathcal P}$ is not a separate panel-root acceleration. The third row of $\mat{P}$ and the first two rows of $\mat{L}$ vanish identically because the deflection $d_3$ is purely out-of-plane. All constant matrices are precomputed, while $\mat{D}_\mathrm{gyro}$, $\mat{K}_d$, and the stiffness-proportional term of $\mat{D}_\mathrm{Ray}$ are evaluated from the instantaneous hub kinematics at each time step.

\noindent The attitude quaternion is advanced with \(\vect{\omega}\) and renormalized after each structural time step. For panel $j$, $\mat K_j=\mat K_{{\rm sp},j}+\mat K_{d,j}$. Let \(m_\mathrm{rb}\) and $\mat{I}_\mathrm{rb}$ denote the mass and inertia matrix of the dry structural configuration, including the hub and the undeformed panel contributions. The external force and torque applied at \(C\) are
\begin{equation}
\vect{F}_\mathrm{ext}=\vect{F}_\mathrm{ctrl}+\vect{F}_\mathrm{sl},
\qquad
\vect{\tau}_\mathrm{ext}=\vect{\tau}_\mathrm{ctrl}+\vect{T}_\mathrm{sl},
\label{eq:structural_external_wrench}
\end{equation}
where the subscript ``ctrl'' denotes commanded actuation. The complete translational--rotational--modal system is
\begin{equation}
\begin{bmatrix}
 m_\mathrm{rb}\mat{I}_3 & \mat{0}
 & \mat{R}_{PB,1}\mat{L}_1
 & \mat{R}_{PB,2}\mat{L}_2 \\
 \mat{0} & \mat{I}_\mathrm{rb}
 & \mat{R}_{PB,1}\mat{P}_1
 & \mat{R}_{PB,2}\mat{P}_2 \\
 \mat{L}_1^T\mat{R}_{PB,1}^T
 & \mat{P}_1^T\mat{R}_{PB,1}^T
 & \mat{M}_1 & \mat{0} \\
 \mat{L}_2^T\mat{R}_{PB,2}^T
 & \mat{P}_2^T\mat{R}_{PB,2}^T
 & \mat{0} & \mat{M}_2
\end{bmatrix}
\begin{bmatrix}
\vect{a}_C \\
\dot{\vect{\omega}} \\
\ddot{\vect{\eta}}_1 \\
\ddot{\vect{\eta}}_2
\end{bmatrix}
=
\begin{bmatrix}
\vect{F}_\mathrm{ext} \\
\vect{\tau}_\mathrm{ext}-\vect{\omega}\times(\mat{I}_\mathrm{rb}\vect{\omega}) \\
-\mat{D}_1\dot{\vect{\eta}}_1-\mat{K}_1\vect{\eta}_1 \\
-\mat{D}_2\dot{\vect{\eta}}_2-\mat{K}_2\vect{\eta}_2
\end{bmatrix}.
\label{eq:coupled_eom_full}
\end{equation}
Here \(\mat{I}_3\) is the \(3\times3\) identity and \(\mat{R}_{PB,j}\) maps vectors from the local frame of panel \(j\) to the body frame. The off-diagonal blocks \(\mat{R}_{PB,j}\mat{L}_j\) and \(\mat{R}_{PB,j}\mat{P}_j\) are, respectively, the translational and rotational inertial couplings derived from \Cref{eq:L_matrix,eq:P_matrix}. This symmetric mass matrix is the full coupling inherited from the panel model of \citet{VallesSanchez2025}.

For the spinning-spacecraft application, the centre-of-gravity translation is constrained, so \(\vect{a}_C=\vect{0}\) and the force row is inactive. The retained rotational--modal subsystem is therefore
\begin{equation}
  \begin{bmatrix}
    \mat{I}_\mathrm{rb}
    & \mat{R}_{PB,1}\mat{P}_1
    & \mat{R}_{PB,2}\mat{P}_2 \\
    \mat{P}_1^T\mat{R}_{PB,1}^T
    & \mat{M}_1
    & \mat{0} \\
    \mat{P}_2^T\mat{R}_{PB,2}^T
    & \mat{0}
    & \mat{M}_2
  \end{bmatrix}
  \begin{bmatrix}
    \dot{\vect{\omega}} \\
    \ddot{\vect{\eta}}_1 \\
    \ddot{\vect{\eta}}_2
  \end{bmatrix}
  =
  \begin{bmatrix}
    \vect{\tau}_\mathrm{ext} - \vect{\omega}\times(\mat{I}_\mathrm{rb}\vect{\omega})
    \\
    -\mat{D}_1\dot{\vect{\eta}}_1 - \mat{K}_1\vect{\eta}_1
    \\
    -\mat{D}_2\dot{\vect{\eta}}_2 - \mat{K}_2\vect{\eta}_2
  \end{bmatrix},
  \label{eq:coupled_eom}
\end{equation}
The panel stiffness and damping matrices in \Cref{eq:coupled_eom_full,eq:coupled_eom} are
\begin{equation}
  \mat{K}_j
  =
  \mat{K}_{\mathrm{sp},j} + \mat{K}_{d,j} \left( \vect{\omega}^{\mathcal{P}_j}, \dot{\vect{\omega}}^{\mathcal{P}_j} \right),
  \qquad
  \mat{D}_j  =
  2\mat{D}_{\mathrm{gyro},j} + \mat{D}_{\mathrm{Ray},j}.
  \label{eq:panel_KD_def}
\end{equation}
For identical panels, $\mat{M}_1=\mat{M}_2=\mat{M}$ and $\mat{K}_{\mathrm{sp},1}=\mat{K}_{\mathrm{sp},2}=\mat{K}_\mathrm{sp}$, while the coupling matrices differ through the panel orientations and attachment locations. \\
The rotational block row of \Cref{eq:coupled_eom} may be written as
\begin{equation}
  \mat{I}_\mathrm{rb}\dot{\vect{\omega}}
  =
  \vect{\tau}_\mathrm{ext} - \vect{\omega}\times(\mat{I}_\mathrm{rb}\vect{\omega}) - \vect{\tau}_\mathrm{flex},
  \label{eq:rigid_flex_balance}
\end{equation}
where the flexible reaction torque is
\begin{equation}
  \vect{\tau}_\mathrm{flex}
  =
  \sum_{j=1}^{2}
  \mat{R}_{PB,j}\mat{P}_j\ddot{\vect{\eta}}_j .
  \label{eq:flex_reaction_torque}
\end{equation}
The corresponding flexible reaction force in the full six-degree-of-freedom model is \(\vect{F}_\mathrm{flex}=\sum_{j=1}^{2}\mat{R}_{PB,j}\mat{L}_j\ddot{\vect{\eta}}_j\). This assembly treats the inertial excitation of the panels and the reaction force and torque returned to the body with the same coupling matrices.
Equivalently, the lower block rows of \Cref{eq:coupled_eom_full} recover the body-coupled modal equations
\begin{equation}
  \mat{M}_j\ddot{\vect{\eta}}_j + \mat{D}_j\dot{\vect{\eta}}_j + \mat{K}_j\vect{\eta}_j
  =
  - \mat{L}_j^T \mat{R}_{PB,j}^T \vect{a}_C
  - \mat{P}_j^T \mat{R}_{PB,j}^T \dot{\vect{\omega}}, \qquad j=1,2.
  \label{eq:panel_body_coupled}
\end{equation}
Thus, the spacecraft translational and angular accelerations excite the flexible modes, while the resulting modal accelerations feed back through \(\vect{F}_\mathrm{flex}\) and \Cref{eq:flex_reaction_torque}. In the constrained-translation spacecraft cases, the first term on the right-hand side of \Cref{eq:panel_body_coupled} vanishes.
The fluid mass and inertia are not included in \(m_\mathrm{rb}\) or $\mat{I}_\mathrm{rb}$; liquid effects enter through the force and torque computed by $\mathcal{F}$. This separation avoids double counting the liquid contribution in the structural balance. After the acceleration solve, the dynamic variables collected in \(\vect{q}_s\) are advanced with a fourth-order Runge--Kutta scheme, and the active translational and angular kinematics are returned as the next Dirichlet iterate.

\section{System energy-balance verification}
\label{sec:energy}

\noindent The partitioned DNS--FSI formulation is verified a posteriori through the mechanical-energy balance of the closed fluid--rigid--flexible system. The diagnostic compares the variation of the fluid and structural mechanical energies with the accumulated external and dissipative works evaluated from the discrete numerical states.

The system comprises a closed tank undergoing translation and rotation, coupled to a rigid structure with flexible appendages. The control volume is the complete tank interior, \(\Omega_f(t)=\Omega_\ell(t)\cup\Omega_g(t)\), fixed in the body frame and moving with the tank in the inertial frame. It is bounded by the tank wall \(\Gamma_w(t)\), with liquid--gas interface \(\Gamma(t)\). No mass crosses \(\Gamma_w\) and no phase change occurs. The reference point \(C\) and frames are defined in \Cref{fig:anna_frames}. All velocity components in the following inner products are expressed in the body frame, although kinetic energy is evaluated from inertial velocities.

The section first defines the stored-energy contributions and power ports used in the post-processing, consistently with the translational--rotational--modal system in \Cref{eq:coupled_eom_full} and the Dirichlet--Neumann coupling introduced in \Cref{sec:coupling}. It then defines the discrete energy-balance residual \(\varepsilon_{\mathrm{bal},h}\), i.e. the unclosed part of the mechanical-energy ledger on grid \(h\). This residual is used in \Cref{sec:convergence_results} as a scalar diagnostic of numerical energy consistency.

\subsection{Stored energies and power ports}
\label{sec:energy_ports}

The total mechanical energy is
\begin{equation}
 E_{\rm tot}=E_{\rm struct}+E_f,\qquad
 E_{\rm struct}=T_{\rm struct}+V_{\rm struct},\qquad
 E_f=T_{f,\rm abs}+V_{g,f}+E_\sigma .
 \label{eq:E_total}
\end{equation}
Here $T_{\rm struct}$ and $T_{f,\rm abs}$ are the absolute kinetic energies of the dry structure and both fluid phases. The potential $V_{\rm struct}$ includes panel deformation and any conservative external loading of the dry structure; $V_{g,f}$ is the fluid gravitational potential, and $E_\sigma$ is the liquid--gas interfacial energy. Conservative external forces are paired with their potentials and excluded from the applied-power port defined below. The dry structure excludes both fluid phases. For the translational--rotational--modal model, define
\begin{equation}
 \vect v_{\rm sys}=[\vect v_C^T,\vect\omega^T,
 \dot{\vect\eta}_1^T,\dot{\vect\eta}_2^T]^T,\qquad
 \mat M_{\rm sys}=\text{mass matrix in \Cref{eq:coupled_eom_full}}.
 \label{eq:Msys_definition}
\end{equation}
The retained structural model directly defines the kinetic energy through its assembled mass matrix:
\begin{align}
 T_{\rm struct}
 &=\tfrac12\vect v_{\rm sys}^T\mat M_{\rm sys}\vect v_{\rm sys}\notag\\
 &=\tfrac12 m_{\rm rb}\|\vect v_C\|^2+
 \tfrac12\vect\omega^T\mat I_{\rm rb}\vect\omega\notag\\
 &\quad+\sum_{j=1}^2(\vect v_C^T\mat R_{PB,j}\mat L_j+
 \vect\omega^T\mat R_{PB,j}\mat P_j)\dot{\vect\eta}_j
 +\tfrac12\sum_{j=1}^2\dot{\vect\eta}_j^T\mat M_j\dot{\vect\eta}_j\notag\\
 &=T_{\rm dry}+T_{\rm hybrid}+T_{\rm flex}.
 \label{eq:T_struct_decomp}
\end{align}
The three contributions are, respectively, the dry-structure rigid-body kinetic energy $T_{\rm dry}$, including translation and rotation, the rigid--flexible kinetic coupling $T_{\rm hybrid}$, and the panel modal kinetic energy $T_{\rm flex}$.
The panel potential energy is
\begin{equation}
 V_{\rm flex}=\tfrac12\sum_{j=1}^2\vect\eta_j^T
 \left(\mat K_{{\rm sp},j}+\mat K_{d,j}\right)\vect\eta_j.
 \label{eq:V_flex}
\end{equation}
Here $\mat K_{{\rm sp},j}$ is the constant linear plate stiffness and $\mat K_{d,j}$ is the rotation-dependent geometric stiffness defined in \Cref{eq:panel_KD_def}. Both matrices are symmetric in the adopted panel model. The time variation of the dynamic stiffness contributes the power
\begin{equation}
 P_K=\tfrac12\sum_{j=1}^2\vect\eta_j^T\dot{\mat K}_{d,j}\vect\eta_j.
 \label{eq:stiffness_variation_power}
\end{equation}
This term accounts for the time variation of $\mat K_{d,j}$, whose contribution is already included in $V_{\rm flex}$.

The absolute fluid velocity, expressed in the body frame, is
\begin{equation}
 \vect u_{\rm abs}=\vect u+\vect U_w,\qquad
 \vect U_w=\vect v_C+\vect\omega\times\vect r_{\rm rel}.
 \label{eq:liquid_abs_velocity}
\end{equation}
Consequently, for either phase $a\in\{\ell,g\}$,
\begin{align}
 T_{a,\rm abs}
 &=\tfrac12\int_{\Omega_a(t)}\rho_a\|\vect u_{\rm abs}\|^2\,dV\notag\\
 &=\tfrac12\int_{\Omega_a(t)}\rho_a\|\vect u\|^2\,dV
 +\int_{\Omega_a(t)}\rho_a\vect u\cdot
 (\vect v_C+\vect\omega\times\vect r_{\rm rel})\,dV\notag\\
 &\quad+\tfrac12\int_{\Omega_a(t)}\rho_a
 \|\vect v_C+\vect\omega\times\vect r_{\rm rel}\|^2\,dV\notag\\
 &=T_{a,\rm rel}+T_{a,\rm cross}+T_{a,\rm frame},
 \qquad T_{f,\rm abs}=\sum_{a\in\{\ell,g\}}T_{a,\rm abs}.
 \label{eq:fluid_energy}
\end{align}
The three contributions are, respectively, the tank-relative kinetic energy $T_{a,\rm rel}$, the signed cross term $T_{a,\rm cross}$, and the rigid-motion contribution $T_{a,\rm frame}$. The latter retains translation, rotation, and their mutual cross term; no constraint on $\vect v_C$ or $\vect\omega$ is imposed in this decomposition.
For uniform gravity, $V_{g,f}=-\sum_a\int_{\Omega_a}\rho_a\vect g^{\mathcal I}\cdot\vect r^{\mathcal I}\,dV$. The liquid--gas interfacial energy retained in the numerical diagnostic is
\begin{equation}
 E_\sigma=\sigma A_\Gamma,
 \label{eq:surface_energy}
\end{equation}
where $A_\Gamma$ is the liquid--gas interfacial area and $\sigma$ is the constant surface tension.

The applied force and torque supply power
\begin{equation}
 P_{\rm ctrl}=\vect v_C^T\vect F_{\rm ctrl}+\vect\omega^T\vect\tau_{\rm ctrl}.
 \label{eq:P_ctrl}
\end{equation}
Any additional non-conservative external loading must be included in this applied-power term. Conservative loading is already included in the stored potentials. The structural damping power is
\begin{equation}
 P_{\rm Ray}=\sum_{j=1}^2\dot{\vect\eta}_j^T\mat D_{{\rm Ray},j}\dot{\vect\eta}_j\geq0,
 \label{eq:P_Ray}
\end{equation}
with any translational damper contribution added when present. The gyroscopic part of $\mat D_j$ in \Cref{eq:panel_damping} is skew-symmetric and therefore contributes no power. The fluid viscous dissipation is
\begin{equation}
 \Phi_{\rm visc}=\sum_{a\in\{\ell,g\}}\int_{\Omega_a}2\mu_a\mat D(\vect u):\mat D(\vect u)\,dV\geq0.
 \label{eq:viscous_dissipation}
\end{equation}
Rigid translation and rotation do not change the symmetric velocity gradient. The fluid load supplies the structural interface power
\begin{equation}
 P_{\rm sl}=\vect v_C^T\vect F_{\rm sl}+\vect\omega^T\vect T_{\rm sl}.
 \label{eq:P_slosh}
\end{equation}
The force and torque act on the tank; the torque is taken about $C$, and all components in the power products are expressed in the body frame. With no slip at the rigid wall, the reciprocal fluid boundary power is $-P_{\rm sl}$.

\subsection{Closed-system energy balance}
\label{sec:closed_energy_balance}

The stored-energy definitions above retain translation and rotation. For the motions considered here---pure translation in the benchmark and rotation about fixed $C$ in the spacecraft case---the subsystem balances are:
\begin{align}
 \frac{dE_{\rm struct}}{dt}&=P_{\rm ctrl}+P_{\rm sl}-P_{\rm Ray}+P_K,
 \label{eq:struct_rate}\\
 \frac{dE_f}{dt}&=-P_{\rm sl}-\Phi_{\rm visc}.
 \label{eq:fluid_rate}
\end{align}
These balances apply to the retained structural approximation. For simultaneous translation and rotation, the body-frame transport terms must also be retained when deriving a balance from \Cref{eq:coupled_eom_full}. Adding \Cref{eq:struct_rate,eq:fluid_rate} cancels the internal wall exchange:
\begin{equation}
 \frac{dE_{\rm tot}}{dt}=P_{\rm ctrl}-P_{\rm Ray}-\Phi_{\rm visc}+P_K.
 \label{eq:global_rate}
\end{equation}
The numerical diagnostic is
\begin{align}
 \varepsilon_{{\rm bal},h}(t)&=E_{{\rm tot},h}(t)-E_{{\rm tot},h}(0)
 -W_{{\rm ctrl},h}(t)\notag\\
 &\quad+W_{{\rm Ray},h}(t)+W_{{\rm visc},h}(t)-W_{K,h}(t),
 \label{eq:ebal_total_system}
\end{align}
where
\begin{equation}
 W_{{\rm ctrl},h}=\int_0^tP_{{\rm ctrl},h}\,d\tau,\quad
 W_{{\rm Ray},h}=\int_0^tP_{{\rm Ray},h}\,d\tau,\quad
 W_{{\rm visc},h}=\int_0^t\Phi_{{\rm visc},h}\,d\tau,\quad
 W_{K,h}=\int_0^tP_{K,h}\,d\tau.
 \label{eq:work_definitions}
\end{equation}
These energy and work definitions are used to assess the energy consistency of the spacecraft simulations in \Cref{sec:convergence_results}.

\section{Reference configuration}
\label{sec:config}

\begin{figure}[H]
\centering
\includegraphics[width=0.75\columnwidth, ]{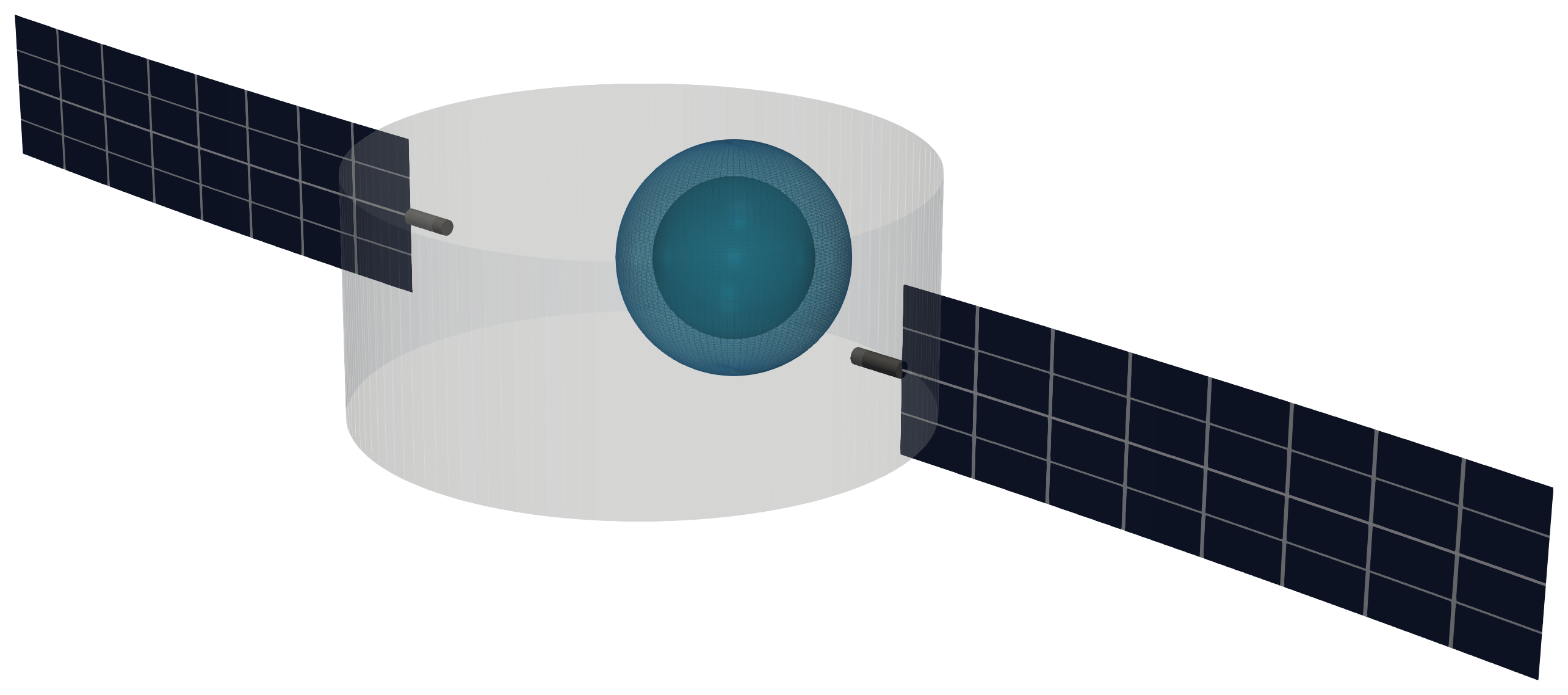}
\caption{Schematic representation of the spinning satellite configuration.}
\label{fig:spacecraft_render}
\end{figure}

The reference case considers the spin-up of a partially filled spherical tank mounted on a rigid–flexible spacecraft under microgravity conditions. As sketched in \Cref{fig:spacecraft_render}, the spacecraft consists of a rigid hub, two identical solar panels mounted on opposite sides of the hub, and a spherical tank placed along the axis perpendicular to the panel direction. 
The equations of motion couple the rigid-body yaw dynamics with the flexible dynamics of the two panels, as defined in \Cref{eq:coupled_eom}. The liquid motion is advanced with the DIVA CFD solver and exchanges forces and torques with the structural model through the partitioned coupling procedure described in \Cref{sec:coupling}.
Table~\ref{tab:params} summarizes the parameters of the reference configuration and the baseline open-loop manoeuvre. 
The position of the tank centre relative to \(C\), expressed in the body frame, is \(\vect{r}_t=[0,L_t,0]^T\), where \(L_t=\SI{0.30}{m}\) is the tank-centre distance.
The body \(z\)-axis is the spin axis, and the tank offset, filling condition, inertial properties, fluid properties, structural parameters, and numerical settings are defined once in \Cref{tab:params}. Unless stated otherwise, that parameter set defines every spacecraft simulation discussed below.
The reference manoeuvre is an open-loop spin-up driven by the control torque \(T_{\mathrm{ctrl},z}\) about the spin axis,
\begin{equation}
    T_{\mathrm{ctrl},z}(t)=
    \begin{cases}
        I_{zz,0}^{\mathrm{tot}}\,\dot{\omega}_{z,\mathrm{ref}}, & 0 \le t < t_{\mathrm{off}},\\
        0, & t \ge t_{\mathrm{off}},
    \end{cases}
    \label{eq:reference_ol_torque}
\end{equation}

Here $I_{zz,0}^{\mathrm{tot}}$ is the reference inertia about the spin axis, $I_{zz,0}^{\mathrm{tot}}=I_{\mathrm{rb},zz}+J_{\ell,z}$, where $J_{\ell,z}=\int_{\Omega_\ell(0)}\rho_\ell\|\vect e_z\times\vect r_{\rm rel}\|^2\,dV$ is the initial liquid inertia about $C$ and $\vect e_z$ is the body-frame spin-axis unit vector; \(\dot{\omega}_{z,\mathrm{ref}}\) is the commanded constant spin acceleration; and \(t_{\mathrm{off}}\) is the torque switch-off time. Their numerical values, including the resulting spin-up torque, are reported in \Cref{tab:params}. For \(t\ge t_{\mathrm{off}}\), the spacecraft evolves freely under the coupled rigid-body, flexible-panel, and liquid-sloshing reactions.

\begin{table}[!ht]
	\centering
	\caption{Parameters of the reference configuration and baseline open-loop spin-up.}
	\label{tab:params}
	\small
	\begin{tabular}{llrl}
		\toprule
		\textbf{Quantity} & \textbf{Symbol} & \textbf{Value} & \textbf{Unit} \\
		\midrule
		\multicolumn{4}{l}{\emph{Spacecraft inertia and panel model}} \\
        Dry structural inertia tensor & $\mathbf{I}_\mathrm{rb}$ & $\mathrm{diag}(8.40,\,8.40,\,0.168)$ & \si{kg.m^2}  \\
        Hub contribution to $I_{zz}$ & $I_{h,zz}$ & 0.159 & \si{kg.m^2} \\
        Panel contribution to $I_{zz}$ & $I_{p,zz}$ & 0.00856 & \si{kg.m^2} \\
		Panel dimensions & $L \times D \times h$ & $1.0\times0.5\times0.001$ & \si{m} \\
        Panel attachment & $x_p$ & $\pm0.30$ & \si{m} \\
        Panel material & $\rho_p,\,E,\,\nu$ & 20.2, 3.95, 0.28 & \si{kg/m^3}, \si{GPa}, -- \\
		Retained modes per panel & $N_m$ & 4 & -- \\
		Panel damping & $\alpha_R,\,\beta_R$ & $10^{-4}$, $1.2\times10^{-3}$ & \si{s^{-1}}, \si{s} \\
        Uncoupled panel frequencies & $f_{1\text{--}4}$ & 0.498, 2.15, 3.12, 7.53 & \si{Hz} \\
		\midrule
		\multicolumn{4}{l}{\emph{CFD setup}} \\
		Tank radius & $R_t$ & 0.05 & \si{m} \\
		Tank centre offset & $\vect{r}_t$ & $[0, 0.30, 0]^T$ & \si{m} \\
		Filling ratio & $\mathrm{FR}$ & 50\% & -- \\
        Liquid mass & $m_\ell$ & 0.364& \si{kg} \\
        Liquid-to-structure inertia ratio & $J_{\ell,z}/I_{\mathrm{rb},zz}$ & 0.198 & -- \\
		Liquid properties & $\rho_\ell,\,\mu_\ell$ & 1410, $10^{-3}$ & \si{kg/m^3}, \si{Pa.s} \\
		Gas properties & $\rho_g,\,\mu_g$ & 2.41, $1.99\times10^{-5}$ & \si{kg/m^3}, \si{Pa.s} \\
		Surface tension & $\sigma$ & 0.0136 & \si{N/m} \\
        Wetting angle / gravity & $\theta_c,\,\mathbf g$ & $0^\circ,\,[0,0,0]$ & \si{deg}, \si{m.s^{-2}} \\
        Grid resolution / domain & $N^3,\,\ell^3$ & $128^3,\,(0.111)^3$ & --, \si{m^3} \\
        Time integrator & $\mathrm{scheme}$ & RK2 & -- \\
        Terminal rate / spin-up duration & $\omega_{z,\mathrm{ref}},\,t_{\mathrm{off}}$ & 1.0, 10 & \si{rad.s^{-1}}, \si{s} \\
        Reference spin acceleration & $\dot{\omega}_{z,\mathrm{ref}}$ & 0.10 & \si{rad.s^{-2}} \\
        Spin-up torque & $T_{\mathrm{ref}}$ & 0.0202 & \si{N.m} \\
		CFL coefficients & $\mathrm{CFL}_\mathrm{conv},\,\mathrm{CFL}_\sigma$ & 0.5, 0.25 & -- \\
		FSI iteration settings & $k_\mathrm{max},\,\varepsilon_\mathrm{FSI},\,\theta^{(0)}$ & 10, $10^{-6}$, 0.7 & -- \\
		\bottomrule
	\end{tabular}
\end{table}

\subsection{Control-oriented LTI representation and controller synthesis}
\label{sec:control}

To synthesize the control laws for the reference configuration, the structural operator \(\mathcal{S}\) and the fluid operator \(\mathcal{F}\) are replaced by LTI representations at their yaw coupling port. The representation of \(\mathcal{S}\) retains the rigid yaw rate and \(N_f=4\) yaw-coupled panel modes, whereas \(\mathcal{F}\) is represented by the acceleration-to-torque map \(\dot{\omega}_z\mapsto\widehat{T}_{\mathrm{sl},z}\) with \(N_s=2\) equivalent-pendulum modes \cite{Abramson1966,Dodge2000}. In this section, a hat denotes a quantity reconstructed or estimated by the reduced model; it is distinct from the raw fixed-point load marked by a hat in \Cref{sec:coupling}. The combined plant state is
\begin{equation}
\vect{x}
=\begin{bmatrix}

\omega_z & \vect{\xi}_s^T & \vect{\nu}_s^T &
\vect{\eta}_f^T & \vect{\nu}_f^T
\end{bmatrix}^{T}\in\mathbb{R}^{13}.
\label{eq:lti_state_vector}
\end{equation}

where \(\omega_z\) is the body yaw rate; \(\vect{\xi}_s=[\xi_{s,1},\ldots,\xi_{s,N_s}]^T\) and \(\vect{\nu}_s=\dot{\vect{\xi}}_s\) are the liquid-surrogate coordinates and velocities; and \(\vect{\eta}_f=[\eta_{f,1},\ldots,\eta_{f,N_f}]^T\) and \(\vect{\nu}_f=\dot{\vect{\eta}}_f\) are the retained panel coordinates and velocities. Each liquid-pendulum mode satisfies
\begin{equation}
\dot \xi_{s,i}=\nu_{s,i},\qquad
\dot \nu_{s,i}=-\omega_{s,i}^{2}\xi_{s,i} -2\zeta_{s,i}\omega_{s,i}\nu_{s,i}+\dot\omega_z,
\label{eq:pendulum_surrogate_state}
\end{equation}
where \(\xi_{s,i}\) is the generalized coordinate of the \(i\)-th pendulum surrogate, \(f_{s,i}\) is the modal frequency, \(\omega_{s,i}=2\pi f_{s,i}\) is the angular frequency, \(\zeta_{s,i}\) is the damping ratio, and \(\dot{\omega}_z\) is the spacecraft angular acceleration driving the surrogate. The reconstructed sloshing torque is
\begin{equation}
\widehat{T}_{\mathrm{sl},z}
=\sum_{i=1}^{N_s}R_i \xi_{s,i}
\label{eq:pendulum_torque_reconstruction}
\end{equation}
where \(R_i\) is the torque weight of mode \(i\). Thus, \(\widehat{T}_{\mathrm{sl},z}\) is the reduced-order prediction of the CFD sloshing torque \(T_{\mathrm{sl},z}^{\mathrm{CFD}}(t) = \vect{e}_z \cdot \vect{T}_\mathrm{sl}(t)\) evaluated in \Cref{sec:diva}.
The identification is performed in two stages. The first stage gives a linear least-squares estimate of the modal torque weights for the spectral seed modes. The second stage uses this estimate to initialize a bounded nonlinear least-squares fit that enforces the physical mass and inertia constraints.
In the first stage, the two dominant peaks of the torque spectrum, separated by at least \(\SI{0.15}{Hz}\), define the frequency seeds. With these frequencies fixed and \(\zeta_{s,i}=0.05\), the seed oscillators are integrated at the \(N_t\) fitting times \(t_k\). Defining the design matrix \(\Phi_{ki}=\xi_{s,i}(t_k)\), the sample vector \(\vect{T}^{\mathrm{CFD}}=[T_{\mathrm{sl},z}^{\mathrm{CFD}}(t_1),\ldots,T_{\mathrm{sl},z}^{\mathrm{CFD}}(t_{N_t})]^T\), and \(\vect{R}=[R_1,\ldots,R_{N_s}]^T\), the modal weights are estimated from
\begin{equation}
\vect{R}^{\star}
= \arg\min_{\vect{R}}
\left\|
\vect{T}^{\mathrm{CFD}} - \mat{\Phi}\vect{R}
\right\|_2^2.
\label{eq:lsq_weight_problem}
\end{equation}
The second stage replaces the unconstrained weights with physically interpretable mass and lever parameters:
\begin{equation}
R_i = s_i\,m_{s,i}\,\lambda_i\,\omega_{s,i}^2,  \qquad
I_{s,i} = \frac{|R_i|}{\omega_{s,i}^{2}} .
\label{eq:weight_mass_coupling}
\end{equation}
where \(s_i\in\{-1,1\}\) defines the torque orientation, \(m_{s,i}\ge 0\) is the participating liquid mass, \(\lambda_i\ge0\) is the equivalent squared lever distance from the spacecraft centre of gravity, and \(I_{s,i}=m_{s,i}\lambda_i\) is the participating rotational inertia. The parameter vector \(\Theta=\{f_{s,i},\zeta_{s,i},m_{s,i},\lambda_i,\xi_{s,i}(t_0),\nu_{s,i}(t_0)\}_{i=1}^{N_s}\), where \(t_0\) is the beginning of the post-spin-up fitting window and the last two entries are the modal initial conditions, is calibrated over the post-spin-up identification window by solving the bounded trust-region least-squares problem
\begin{equation}
\Theta^\star=\mathop{\mathrm{arg\,min}}_{\Theta}
\left\{
\sum_{k=1}^{N_t}\left[
T_{\mathrm{sl},z}^{\mathrm{CFD}}(t_k)
-\sum_{i=1}^{N_s}R_i\xi_{s,i}(t_k;\Theta)
\right]^2+J_\mathrm{sep}+J_\mathrm{seed}
\right\}.
\label{eq:physical_pendulum_objective}
\end{equation}
Here \(\Theta^\star\) is the fitted parameter set. The penalty \(J_\mathrm{sep} = w_\mathrm{sep}^2 N_t \sigma_T^2 \left[\max\left(0,\, 1 - |f_{s,2}-f_{s,1}|/\SI{0.1}{Hz}\right)\right]^2\) enforces a minimum frequency separation of \(\SI{0.1}{Hz}\), while \(J_\mathrm{seed} = w_\mathrm{seed}^2 N_t \sigma_T^2 \sum_{i=1}^{N_s} \left[(f_{s,i}-f_{s,i}^\mathrm{seed})/(0.05 f_{s,i}^\mathrm{seed})\right]^2\) penalizes displacement from the spectral seeds. Their residual weights are \(w_\mathrm{sep}=5\) and \(w_\mathrm{seed}=0.03\), respectively, after scaling by the torque standard deviation \(\sigma_T\) and the number of timesteps \(N_t\). Each frequency is bounded within \(5\%\) of its seed, and the remaining constraints are \(10^{-4}\leq\zeta_{s,i}\leq0.05\), \(\sum_{i=1}^{N_s} m_{s,i}\leq m_\ell\), and \(0\leq\lambda_i\leq(L_t+R_t)^2\),
where \(m_\ell\) is the total liquid mass, \(L_t\) is the tank-centre distance defined in \Cref{sec:config}, and \(R_t\) is the tank radius. Thus, \(L_t+R_t\) is the maximum distance from \(C\) to the spherical tank wall. The torque is fitted because it is the Neumann load through which \(\mathcal{F}\) enters the yaw dynamics. The imposed spin-up transient is excluded from the identification window.

The structural LTI block uses the zero-spin yaw projection of the panel model. The matrices \(\mat{M}_f\), \(\mat{K}_f=\mat{K}_\mathrm{sp}\), and \(\mat{D}_f=\mat{D}_\mathrm{Ray}\) are the retained modal mass, stiffness, and damping matrices, and \(\vect{p}_z\) is the yaw projection of the two-panel angular-acceleration coupling. The resulting second-order assembly is
\begin{equation}
\mat{M}_g
\begin{bmatrix}
\dot\omega_z\\
\dot{\vect{\nu}}_f
\end{bmatrix}
=
\begin{bmatrix}
u+\widehat{T}_{\mathrm{sl},z}\\
-\mat{K}_f\vect{\eta}_f-\mat{D}_f\vect{\nu}_f
\end{bmatrix},
\qquad
\mat{M}_g=
\begin{bmatrix}
I_\mathrm{base} & \vect{p}_z^T\\
\vect{p}_z & \mat{M}_f
\end{bmatrix},
\label{eq:control_second_order_yaw_flex}
\end{equation}
with
\begin{equation}
I_\mathrm{base}=I_{\mathrm{rb},zz}
+I_{\ell,\mathrm{fix}}+\sum_i I_{s,i}.
\label{eq:control_base_inertia}
\end{equation}
Here \(\mat{M}_g\) is the generalized yaw--flexible mass matrix, \(u=T_{\mathrm{ctrl},z}\) is the commanded yaw torque, \(I_{\mathrm{rb},zz}\) is the dry yaw inertia, and \(I_{\ell,\mathrm{fix}}\) is the inertia of the liquid not assigned to the moving modes. Substitution of the pendulum equations and inversion of \(\mat{M}_g\) give the state-space representation for the 13-state vector \(\vect{x}\) defined in \Cref{eq:lti_state_vector}:
\begin{equation}
\dot{\vect{x}}=\mat{A}\vect{x}+\vect{B}u,
\qquad
\vect{y}=\mat{C}\vect{x}
=\begin{bmatrix}\omega_z&\widehat{T}_{\mathrm{sl},z}\end{bmatrix}^{T}.
\label{eq:control_lti_state}
\end{equation}
where \(\mat{A}\in\mathbb{R}^{13\times13}\) is the state matrix, \(\vect{B}\in\mathbb{R}^{13}\) is the torque-input vector, \(\vect{y}\in\mathbb{R}^{2}\) is the output vector, and \(\mat{C}\in\mathbb{R}^{2\times13}\) is the output matrix. The scalar feedback measurement is \(y_m=\mat{C}_m\vect{x}=\omega_z\), where \(\mat{C}_m\) is the first row of \(\mat{C}\); \(\widehat{T}_{\mathrm{sl},z}\) is retained as a diagnostic output. The explicit block matrices are given in \Cref{app:control_lti_matrices}.

The control study retains only an LQG servo. Let \(\omega_r(t)\) be the commanded yaw-rate trajectory and let \(z\) be the integral of the yaw-rate tracking error. The augmented controller is
\begin{equation}
\dot z=\omega_z-\omega_r,\qquad
u_\mathrm{LQG}=-K_i z-\vect{K}_x\widehat{\vect{x}}.
\label{eq:lqg_control_law}
\end{equation}
Here \(K_i\) is the integral gain, \(\vect{K}_x\) is the state-feedback gain, and \(\widehat{\vect{x}}\) is the observer estimate of the plant state \(\vect{x}\) defined in \Cref{eq:lti_state_vector}. The gain \([K_i\ \vect{K}_x]\) is obtained from the continuous-time algebraic Riccati equation for the augmented state \(\vect{x}_a=[z\ \vect{x}^{T}]^{T}\), using diagonal state weights \(Q_z=Q_\omega=200\), \(Q_s=20\), \(Q_f=5\), and actuator weight \(R_u=2\). A steady Kalman--Bucy observer reconstructs the state estimate \(\widehat{\vect{x}}\) of \(\vect{x}\) from the measured output \(y_m=\omega_z\), using process covariance \(\mat{Q}_\mathrm{KF}=5\times10^{-4}\mat{I}_{13}\) and measurement covariance \(R_\mathrm{KF}=10^{-3}\). Its continuous dynamics are integrated over each accepted CFD time step with RK4 substeps no larger than \(\SI{2e-3}{s}\). The yaw-axis command is limited to \(\lvert u\rvert\leq\SI{0.35}{N.m}\), and the integral state is held while the command is saturated. The replay peak is \(\SI{2.20e-2}{N.m}\), so saturation is not activated in the reported manoeuvre.

\section{Results and discussion}
\label{sec:results}

The results are presented in the same order as the modelling claims. \Cref{sec:validation} first assesses the translational force--motion exchange. The numerical consistency of the coupled computations is then assessed through grid convergence of the energy balance in \Cref{sec:convergence_results}. \Cref{sec:pm_ol_comparison} compares matched prescribed-motion, rigid-FSI, and flexible-FSI simulations and tests how the identification architecture affects reduced-model transfer. Finally, \Cref{sec:cl_campaign} replays the LQG controller in the nonlinear flexible CFD--FSI plant.

\subsection{Coupling validation}
\label{sec:validation}
\begin{figure}[H]
\centering
\includegraphics[width=0.75\columnwidth]{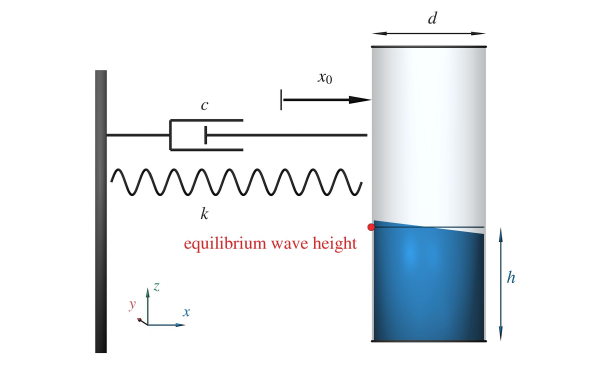}
\caption{Schematic representation of the validation configuration. The partially filled cylindrical tank translates along $x$ from the imposed initial displacement $x_0$, while the free-surface elevation is monitored at the left wall relative to its equilibrium level.}
\label{fig:smd_render}
\end{figure}

\noindent To validate the implementation of the fixed-point coupling between the DIVA CFD solver and the rigid–flexible structural model, numerical results are compared against experimental data available in the literature. The reference case is the experiment of Peterson et al. \cite{peterson1989nonlinear}, which investigates nonlinear liquid sloshing coupled with the translational dynamics of a spacecraft-like system. The experimental setup consists of a cylindrical container partially filled with liquid coupled with a spring--mass--damper (SMD) system. The cylinder is free to translate in the horizontal direction under the combined action of the structural restoring and damping forces and the hydrodynamic loads generated by the sloshing liquid, as illustrated in \Cref{fig:smd_render}.

\noindent The cylinder has radius $a=d/2=0.0155\, m$ and is filled up to a liquid height $h=d$. The liquid considered consists of a water solution with $2\%$ photoflo. \\
The liquid and gas densities are set to $\rho_\ell=1000 \, \si{kg/m^3}$ and $\rho_g=1.3 \, \si{kg/m^3}$, respectively, while the corresponding dynamic viscosities are $\mu_\ell=10^{-3} \, \si{Pa.s}$ and $\mu_g=1.3\times10^{-6} \, \si{Pa.s}$.
The validation case considered here corresponds to the free-decay response of the cylinder displacement in the $x$-direction with Bond number $Bo=\rho_\ell g a^2/\sigma=66$, based on the cylinder radius $a$, mass ratio $\hat{m} = m_f/m_{cyl} = 0.16$, frequency ratio $\hat{f} = \omega_{s,1}/\omega_0 = \omega_{s,1}/\left( \sqrt{k/m_{cyl}}\right)$ with $\omega_{s,1}$ from \cite{peterson1989nonlinear}, damping ratio $\zeta = 9.5 \%$, and excitation level \(\Xi_{ex}=0.013\). In the numerical free-decay problem, this excitation is represented by releasing the cylinder from the corresponding non-equilibrium position.
These parameters give a liquid surface tension \(\sigma_l=\rho_l g a^2/Bo=\SI{0.0357}{N/m}\), a dry cylinder mass \(m_{\mathrm{cyl}}=\SI{0.146}{kg}\), a spring stiffness \(k=\SI{149}{N/m}\), a damping coefficient \(c=\SI{0.893}{kg/s}\), and an initial displacement \(x_0=\SI{6.05e-4}{m}\).
Photoflo is a surfactant used in the experiments to reduce surface tension and contact-angle hysteresis at the liquid--gas--solid contact line. Peterson et al. report that the \(2\%\) solution reduced both the fluid--container contact angle and contact-angle hysteresis to approximately zero. The resulting low-hysteresis behaviour is represented in the numerical model by neglecting contact-line dissipation. Imposing a \(0^\circ\) contact angle would represent complete wetting, leading to the formation of a thin liquid film all over the solid boundaries. Because this normal-gravity benchmark focuses on the macroscopic free-surface response, the wall-film state is not included, and a fixed \(90^\circ\) contact angle is instead used to define the interface geometry. The results below show that this approximation is sufficient to reproduce the cylinder displacement and the main oscillation dynamics considered in the validation.
In the following, time is normalized by the dry spring--mass period
\begin{equation}
  T_\mathrm{sm}=2\pi\sqrt{\frac{m_\mathrm{cyl}}{k}},
  \label{eq:validation_Tsm}
\end{equation}
which is the period of the dry cylinder oscillator. The frozen-liquid SMD comparison is computed separately from $(m_{\rm cyl}+m_f)\ddot x+c\dot x+kx=0$, with the same release displacement and zero initial velocity, the liquid contributes only as a  rigidly attached mass.

\begin{figure}[!htbp]
\centering
\includegraphics[width=0.49\textwidth]{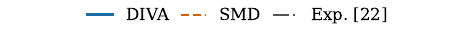}\par\nointerlineskip\vspace{1pt}
\begin{subfigure}[t]{0.49\textwidth}
\centering
\includegraphics[width=\linewidth]{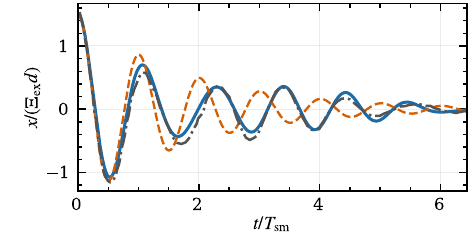}
\caption{Normalized cylinder displacement.}
\label{fig:cylinder_nondimensional_displacement}
\end{subfigure}
\hfill
\begin{subfigure}[t]{0.49\textwidth}
\centering
\includegraphics[width=\linewidth]{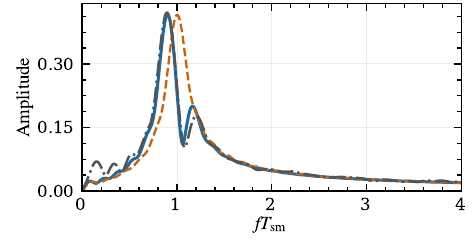}
\caption{Normalized displacement spectrum.}
\label{fig:cylinder_nondimensional_fft}
\end{subfigure}
\par\nointerlineskip\vspace{4pt}
\caption{Validation of the coupled cylinder response against the Peterson free-decay experiment. The displacement comparison includes the fully coupled
\textsc{DIVA} prediction as the blue line, the experimental reference in Peterson \cite{peterson1989nonlinear} in grey, and the frozen-liquid spring-mass response used to isolate the frequency shift induced by sloshing in orange.}
\label{fig:cylinder_nondimensional}
\end{figure}

\begin{figure}[!tbp]
\centering
\includegraphics[width=\textwidth]{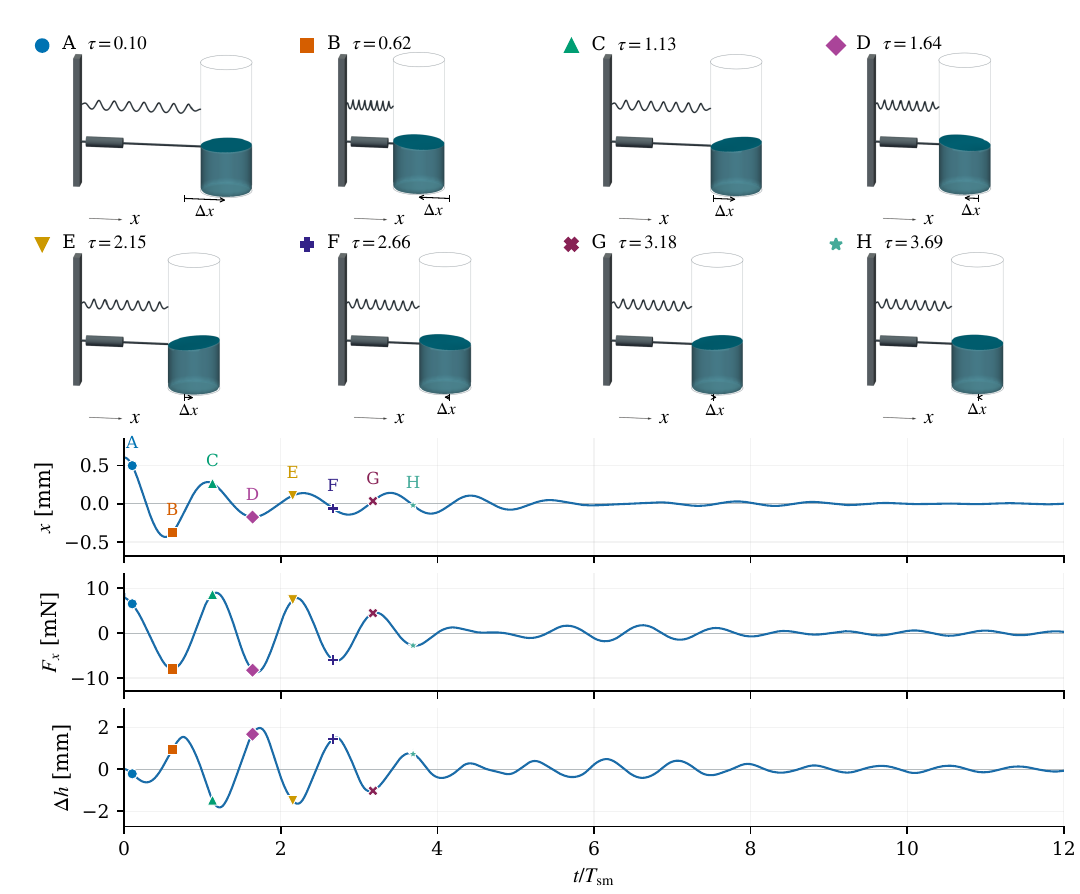}
\caption{Coupled cylinder motion and liquid response ($\sigma=\SI{0.0357}{N.m^{-1}}$). Snapshots A--H correspond to the marked instants $\tau=t/T_{\mathrm{sm}}$ on the curves. From top to bottom, the histories show tank displacement, hydrodynamic force, and relative left-wall wave height. The displacement arrows indicate $\Delta x=x-x_{\mathrm{eq}}$.}
\label{fig:cylinder_slosh}
\end{figure}

\noindent \Cref{fig:cylinder_nondimensional} compares the experimental free-decay response with the numerical prediction in terms of normalized cylinder displacement and dominant oscillation frequency. The \textsc{DIVA}-based coupled simulation reproduces the oscillation frequency of the rigid–flexible–liquid system and remains in good phase agreement with the experimental measurements over the time interval considered. The frozen-liquid response, in which the liquid contributes only as an additional rigid mass attached to the cylinder, provides a reference for interpreting the frequency reduction induced by liquid motion. This confirms that the solver captures the effective softening introduced by the sloshing dynamics.
After release, the liquid motion generates hydrodynamic forces that are generally opposed to the instantaneous motion of the container, as shown in \Cref{fig:cylinder_slosh} (middle curve). This behaviour is consistent with the lag of part of the liquid relative to the cylinder motion, which is also visible in the evolution of the free-surface elevation near the left wall of the cylinder reported in \Cref{fig:cylinder_slosh} (bottom curve). The resulting fluid forces are nonlinear and depend on both the instantaneous interface shape and the motion history of the container. These results highlight the strongly coupled nature of the problem and demonstrate the ability of the fixed-point coupling strategy to reproduce the main features of the experimental fluid–structure response.

\subsection{Grid convergence of the energy balance}
\label{sec:convergence_results}

The energy balance defined in \Cref{sec:energy} is applied here to the spacecraft configuration. For these simulations, $\vect v_C=\vect0$ and $\vect g=\vect0$. No additional conservative structural potential is applied, so $V_{\rm struct}=V_{\rm flex}$ and $V_{g,f}=0$. The rigid and coupling kinetic energies and the power ports reduce to
\begin{align}
 \tfrac12\vect\omega^T\mat I_{\rm rb}\vect\omega&=T_{\rm dry},\notag\\
 \sum_{j=1}^2\vect\omega^T\mat R_{PB,j}\mat P_j\dot{\vect\eta}_j&=T_{\rm hybrid},\notag\\
 \vect\omega^T\vect\tau_{\rm ctrl}&=P_{\rm ctrl},\qquad
 \vect\omega^T\vect T_{\rm sl}=P_{\rm sl}.
 \label{eq:spacecraft_energy_ports}
\end{align}
The total stored energy is therefore
\begin{align}
 E_{\rm tot}&=T_{\rm dry}+T_{\rm hybrid}+T_{\rm flex}+V_{\rm flex}\notag\\
 &\quad+T_{\ell,\rm abs}+T_{g,\rm abs}+\sigma A_\Gamma.
 \label{eq:spacecraft_total_energy}
\end{align}
For this study case, gas-phase kinetic energy and viscous dissipation are considered negligible compared with their liquid-phase counterparts for the present property ratios, \(\rho_g/\rho_\ell\simeq1.71\times10^{-3}\) and \(\mu_g/\mu_\ell\simeq1.99\times10^{-2}\), and are therefore neglected in the balance. The plots group the liquid kinetic and interfacial energies as $E_\ell=T_{\ell,\rm abs}+\sigma A_\Gamma$. Since $\vect v_C=\vect0$, $T_{\ell,\rm frame}$ reduces to the rotational contribution $T_{\ell,\rm rot}$. These definitions apply to all spatial refinements.

The numerical assessment uses the flexible reference configuration presented in \cref{sec:config} over the common interval \(0\leq t\leq\SI{60}{s}\). The baseline discretization employs a $64^3$ Cartesian grid with convective and capillary stability coefficients $\mathrm{CFL}_{\mathrm{conv}}=0.5$ and $\mathrm{CFL}_{\sigma}=0.25$, respectively. 

\begin{figure}[!htbp]
\centering
\begin{subfigure}[t]{0.49\textwidth}
\vspace{0pt}
\centering
\includegraphics[width=\linewidth]{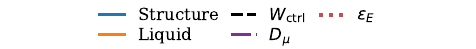}\par\nointerlineskip\vspace{1pt}
\includegraphics[width=\linewidth]{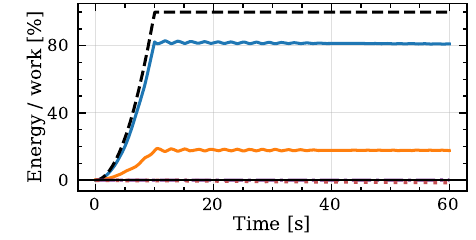}
\caption{Stored energies and balance.}
\end{subfigure}
\hfill
\begin{subfigure}[t]{0.49\textwidth}
\vspace{0pt}
\centering
\includegraphics[width=\linewidth]{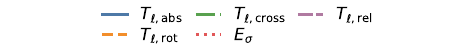}\par\nointerlineskip\vspace{1pt}
\includegraphics[width=\linewidth]{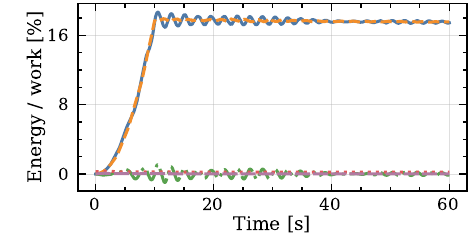}
\caption{Liquid-energy decomposition.}
\end{subfigure}
\par\nointerlineskip\vspace{4pt}
\begin{subfigure}[t]{0.49\textwidth}
\vspace{0pt}
\centering
\includegraphics[width=\linewidth]{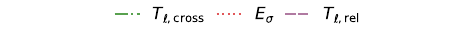}\par\nointerlineskip\vspace{1pt}
\includegraphics[width=\linewidth]{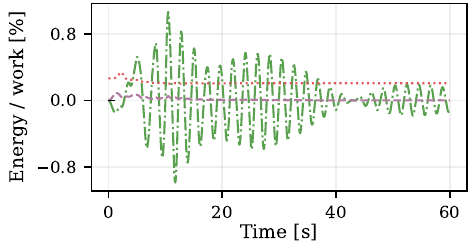}
\caption{Smaller liquid-energy terms.}
\end{subfigure}
\hfill
\begin{subfigure}[t]{0.49\textwidth}
\vspace{0pt}
\centering
\includegraphics[width=\linewidth]{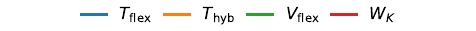}\par\nointerlineskip\vspace{1pt}
\includegraphics[width=\linewidth]{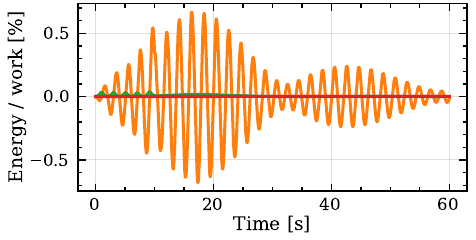}
\caption{Flexible stored energies and stiffness-variation work.}
\end{subfigure}
\par\nointerlineskip\vspace{4pt}
\caption{Energy partition of the nominal-CFL \(64^3\) flexible-FSI case with balance diagnostics, normalized by the control-work scale. (a) Structural and liquid stored energies, supplied control work, cumulative viscous dissipation, and signed energy discrepancy. (b) Absolute liquid kinetic energy, its relative, cross, and rotational contributions, and surface energy; (c) resolves the three smaller terms. (d) Flexible kinetic energy, rigid--flexible kinetic coupling, panel potential energy, and stiffness-variation work \(W_K\).}
\label{fig:energy_dissipation_discrepancy}
\end{figure}

\Cref{fig:energy_dissipation_discrepancy} presents the energy partition of the baseline case, with nominal-CFL and \(64^3\) grid, using the definitions introduced in \Cref{sec:energy_ports}. The structural contribution is approximately \(82.1\%\) of the total input-work scale, while the absolute liquid kinetic energy accounts for \(17.7\%\). The small remainder is primarily surface energy; the direct flexible kinetic and strain-energy contributions are negligible on this scale. At \(t=\SI{60}{s}\), the signed energy-balance discrepancy is \(-1.35\%\). The liquid-energy decomposition in panel (b), defined in \Cref{eq:fluid_energy}, shows that the rotational contribution dominates the absolute liquid kinetic energy, while the signed cross term ranges from \(-0.987\%\) to \(1.07\%\) and changes sign over the interval. The relative liquid kinetic contribution is small, consistent with limited tank-relative liquid motion, and the surface energy ranges from \(0.203\%\) to \(0.341\%\). The undeformed panels are included in the dry-structure inertia, whereas the deformation-related kinetic, rigid--flexible coupling, and panel potential energies are represented separately by $T_\mathrm{flex}$, $T_\mathrm{hybrid}$, and $V_\mathrm{flex}$, respectively. The coupling term $T_\mathrm{hybrid}$ oscillates about zero.

The consistency of this energy accounting is assessed through the discrete closed-system energy-balance residual \(\varepsilon_{\mathrm{bal},h}\) defined in \Cref{eq:ebal_total_system}. Spatial sensitivity is evaluated at the nominal CFL setting using the $64^3$, $128^3$, and $256^3$ grids. The three simulations use the physical parameters, coupling tolerance, and torque command reported in \cref{tab:params}.

\begin{figure}[!htbp]
\centering
\includegraphics[width=\linewidth]{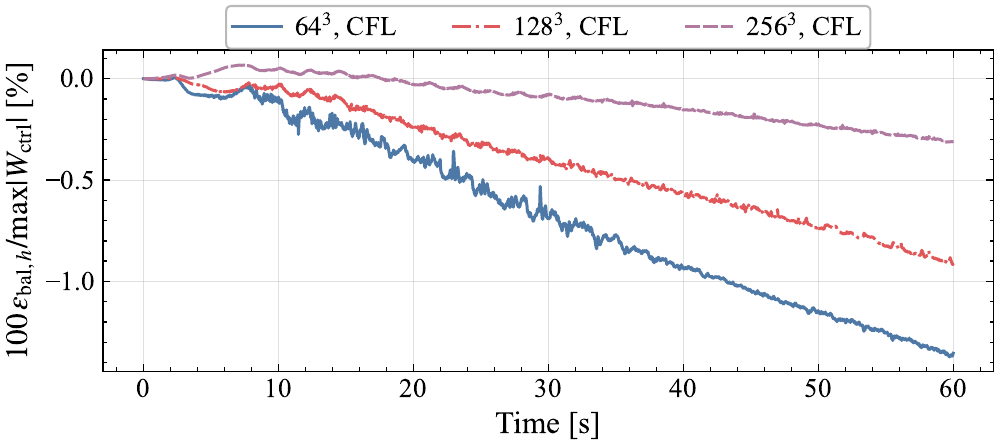}
\caption{Normalized closed-system energy residual for the three grids at nominal CFL over the common \(\SI{60}{s}\) interval. Values are percentages of the maximum absolute cumulative control work.}
\label{fig:updated_energy_convergence}
\end{figure}

\begin{figure}[!htbp]
\centering
\includegraphics[width=\linewidth]{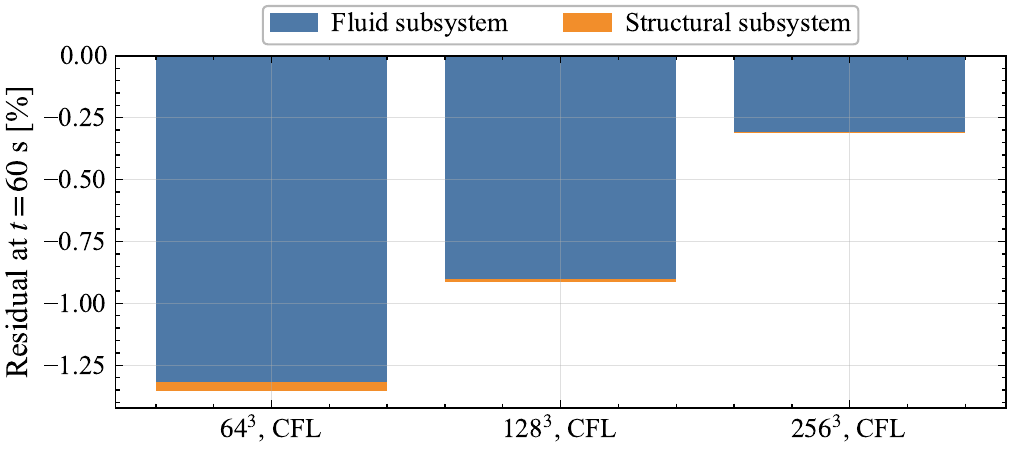}
\caption{Fluid and structural contributions to the normalized energy residual at \(t=\SI{60}{s}\) for the three grids at nominal CFL.}
\label{fig:updated_subsystem_energy}
\end{figure}

Spatial refinement reduces the residual magnitude monotonically from $1.35\%$ on the $64^3$ grid to $0.914\%$ on the $128^3$ grid and $0.312\%$ on the $256^3$ grid. This behavior indicates that the dominant contribution arises from continuum-to-discrete consistency errors in the fluid formulation, which are progressively reduced as the spatial representation is refined.
Due to numerical diffusion during level-set advection and redistancing, minor liquid-volume losses occur over time. At \(t=\SI{60}{s}\), the cumulative loss relative to the initialized state decreases from \(0.00431\%\) on the \(64^3\) grid to \(0.00203\%\) on the \(128^3\) grid and \(0.000507\%\) on the \(256^3\) grid at nominal CFL. These losses remain small throughout the manoeuvre.

The subsystem decomposition in \Cref{fig:updated_subsystem_energy} confirms that the discrepancy is predominantly associated with the fluid subsystem. For the baseline $64^3$ case, the total residual of $-1.35\%$ comprises $-1.32\%$ from the fluid and $-0.0367\%$ from the structure. At $256^3$, the corresponding contributions decrease to $-0.307\%$ and $-0.00501\%$. All accepted steps satisfy the prescribed Dirichlet--Neumann convergence criterion, excluding failed coupling iterations as the source of the observed trend.

\begin{figure}[!htbp]
\centering
\includegraphics[width=\linewidth]{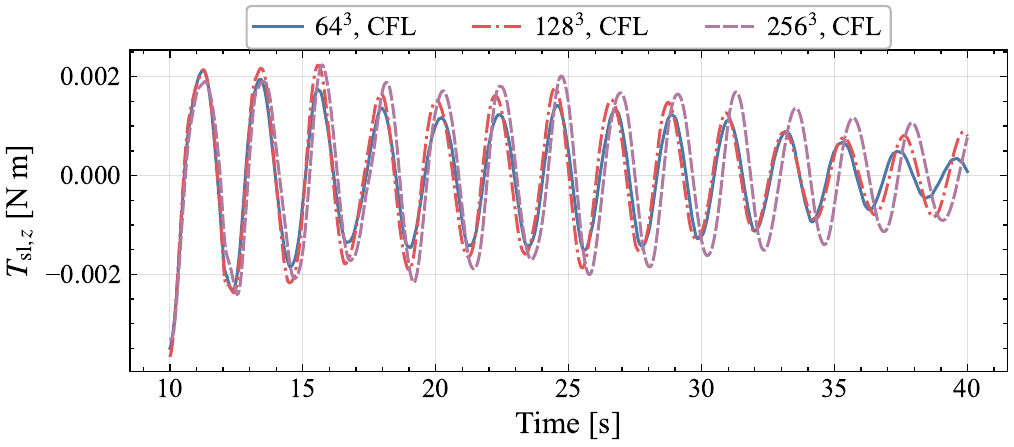}
\caption{Spatial sensitivity of the sloshing torque $T_{\mathrm{sl},z}(t)$ about $C$ over \(10\leq t\leq\SI{40}{s}\). The baseline $64^3$ grid is compared with the $128^3$ and $256^3$ grids at nominal CFL, with the corresponding CFL-limited time steps.}
\label{fig:updated_dynamic_convergence}
\end{figure}

\Cref{fig:updated_dynamic_convergence} shows the spatial-refinement curves. Increasing the grid resolution reduces the numerical damping and preserves richer oscillations, so that the $128^3$ and the $256^3$ cases retain a more persistent sloshing response than the baseline grid. This improvement, however, comes with an additional computational cost in the coupled simulations: each physical time step must repeat the CFD advance until the Dirichlet--Neumann fixed point is satisfied, and the Aitken-relaxed iterations still require about four to five additional CFD resolves per step. Within the present campaign, the $128^3$ grid is consequently retained for the matched PM, rigid-FSI, and flexible-FSI calculations as the compromise between late-time signal retention and computational feasibility.

\subsection{Prescribed-motion, rigid open-loop, and flexible open-loop comparisons}
\label{sec:pm_ol_comparison}

The reference manoeuvre is simulated with prescribed tank motion (PM), rigid open-loop coupling, and flexible open-loop coupling. The three configurations use the same liquid, geometry, \(128^3\) grid, and spin-up command. They differ only in the treatment of the spacecraft dynamics and therefore provide a direct test of the effect of fluid--structure feedback on the liquid response.
The prescribed-motion configuration represents the standard CFD approach and corresponds to the original DIVA formulation, previously validated against FLUIDICS microgravity experiments. The nominal tank kinematics are imposed in the fluid subsystem \(\mathcal{F}\) through the volumetric non-inertial forcing \(\vect{f}_{\mathrm{vol}}\) of \Cref{eq:non_inertial_force}. The liquid loads are evaluated, but they do not modify the prescribed motion. In the rigid open-loop calculation, the same spin-up is applied as \(T_{\mathrm{ctrl},z}=\SI{2.02e-2}{N.m}\) during the first \(\SI{10}{s}\), and the fluid and rigid-body operators exchange Neumann and Dirichlet data through the coupling procedure of \Cref{sec:coupling}. The flexible calculation retains this exchange and adds the four panel modes of \Cref{sec:flex_model}. Thus, PM describes the liquid under imposed kinematics, whereas the two FSI calculations allow the liquid loads to alter the spacecraft motion.

\begin{figure}[!htbp]
\centering
\includegraphics[width=\textwidth]{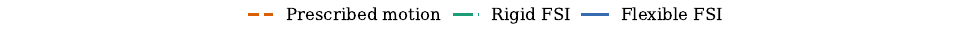}\par\nointerlineskip\vspace{1pt}
\begin{subfigure}[t]{\textwidth}
\centering
\includegraphics[width=\linewidth]{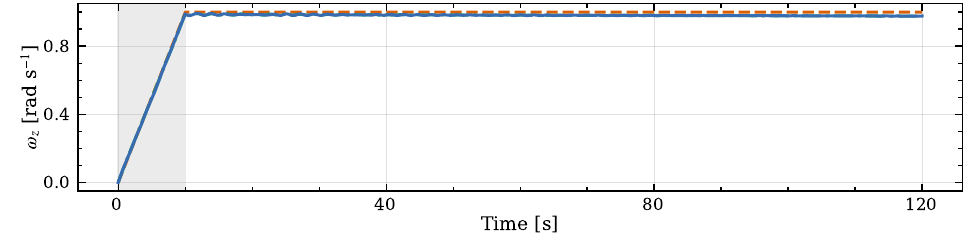}
\caption{Yaw rate.}
\end{subfigure}
\par\nointerlineskip\vspace{4pt}
\begin{subfigure}[t]{\textwidth}
\centering
\includegraphics[width=\linewidth]{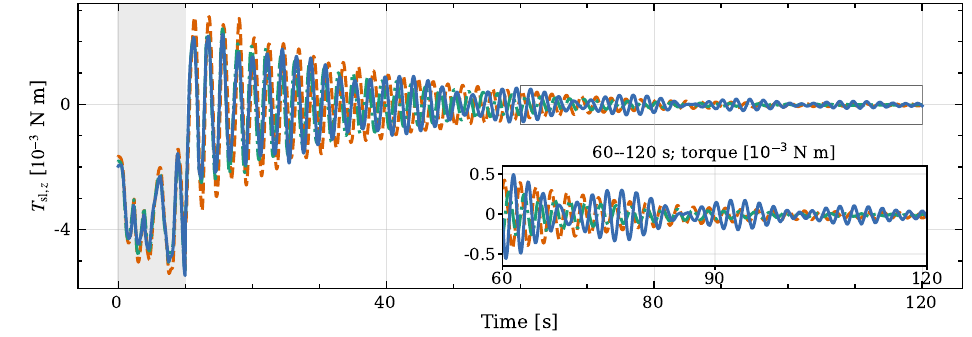}
\caption{Sloshing torque with late-time zoom.}
\end{subfigure}
\par\nointerlineskip\vspace{4pt}
\begin{subfigure}[t]{\textwidth}
\centering
\includegraphics[width=\linewidth]{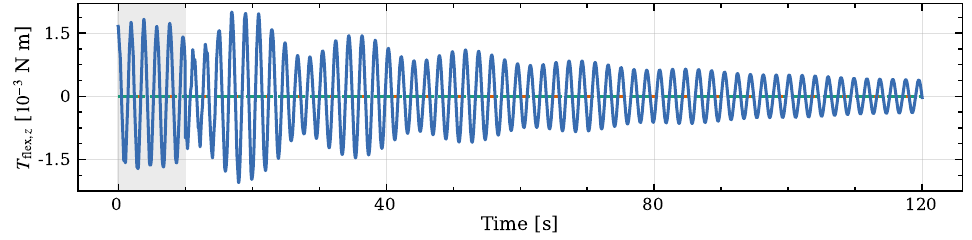}
\caption{Flexible reaction torque.}
\end{subfigure}
\par\nointerlineskip\vspace{4pt}
\caption{Comparison of prescribed-motion, rigid-FSI, and flexible-FSI simulations for the reference manoeuvre. The panels show (a) yaw rate, (b) sloshing torque, (c) the flexible reaction torque. The shaded interval denotes the spin-up. The inset in (b) magnifies the boxed interval, $60\leq t\leq120$ s; its torque scale is shown explicitly.}
\label{fig:framework_comparison}
\end{figure}

\Cref{fig:framework_comparison} shows the yaw rate and the torques about the spin axis for the three configurations. The angular velocity of the PM simulation represents the reference value. Open-loop coupled simulations show a gradual departure. This phenomenon is linked to the kinetic energy lost due to the inconsistencies between the spatial numerical discretization of the fluid and the calculation of the continuous kinetic energy. In the case of prescribed motion, the sloshing torques have greater oscillations than in the cases coupled with the structure. During the early coast (\(10\leq t\leq\SI{60}{s}\)), the mean-removed sloshing-torque RMS is \(1.21\times10^{-3}\), \(9.79\times10^{-4}\), and \(9.56\times10^{-4}\,\si{N.m}\) for PM, rigid FSI, and flexible FSI. The PM value is \(23.6\%\) larger than the rigid-FSI value. During the final coast, \(60\leq t\leq\SI{120}{s}\), the flexible case retains a sloshing-torque RMS of \(1.47\times10^{-4}\,\si{N.m}\), compared with \(8.11\times10^{-5}\,\si{N.m}\) for rigid FSI, an increase of \(81.5\%\). The flexible reaction torque is comparable to the liquid torque during the initial response and remains significant during the late coast, where it changes the persistence of the coupled response.

\begin{figure}[!htbp]
\centering
\includegraphics[width=\textwidth]{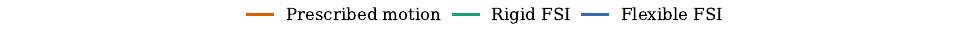}\par\nointerlineskip\vspace{1pt}
\includegraphics[width=\linewidth]{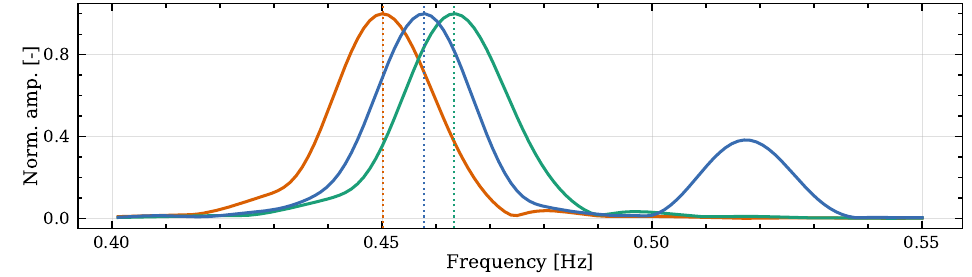}
\caption{Spectra of the sloshing torque over the complete post-spin interval $t\ge10s$. Dotted lines mark the dominant response peaks for PM, rigid FSI, and flexible FSI.}
\label{fig:spectral_peak_resolution}
\end{figure}

The spectra in \Cref{fig:spectral_peak_resolution} use uniformly resampled torque histories over \(10\)--\(\SI{120}{s}\), with linear detrending, a Hann window, and zero padding. The spectra of the sloshing z torque in \Cref{fig:spectral_peak_resolution} locate the dominant liquid response at \(0.450\), \(0.463\), and \(\SI{0.458}{Hz}\) for PM, rigid FSI, and flexible FSI. The flexible simulation also contains a distinct frequency content near \(\SI{0.518}{Hz}\), outside the liquid band.

\begin{figure}[!htbp]
\centering
\includegraphics[width=\textwidth]{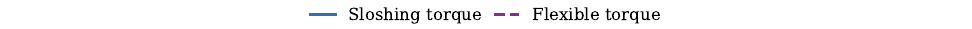}\par\nointerlineskip\vspace{1pt}
\includegraphics[width=\linewidth]{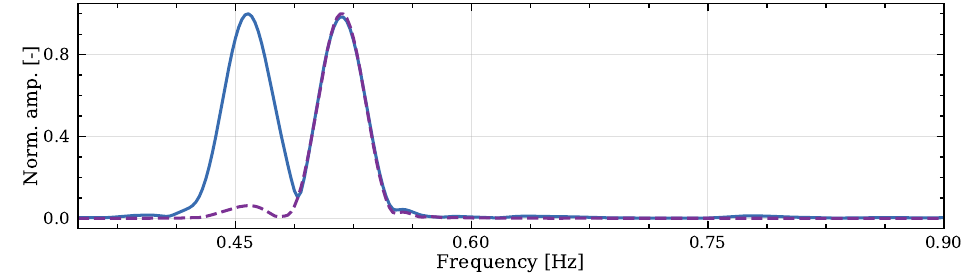}
\caption{Late-time sloshing- and panel-reaction-torque spectra for flexible FSI over \(60\leq t\leq\SI{120}{s}\).}
\label{fig:flexible_interaction}
\end{figure}

The \(\SI{0.518}{Hz}\) component is present in both the sloshing and panel-reaction spectra in \Cref{fig:flexible_interaction}. Its normalized amplitude is \(0.984\), whereas it remains negligible in PM and rigid FSI.

The three configurations therefore present a variation in the oscillation frequency of the liquid depending on the system considered. For this configuration, where the structural and sloshing frequencies are close to each other, the post-manoeuvre liquid dynamics are altered by the structural feedback. Accurately capturing the coupled sloshing dynamics is therefore crucial for attitude control of spacecraft with large flexible appendages and liquid propellant. To demonstrate this, we use the two-pendulum ROM liquid representation explained in \Cref{sec:control}. To quantify this modelling consequence, this two-mode liquid representation is identified separately from the PM, rigid-FSI, and flexible-FSI records. Each identified model is then inserted without retuning into the same rigid--flexible LTI plant. This common-plant replay keeps the structural realization fixed and isolates the effect of the liquid model obtained from each simulation architecture.

\begin{figure}[!htbp]
\centering
\includegraphics[width=0.49\textwidth]{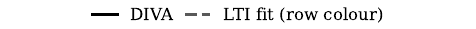}\par\nointerlineskip\vspace{1pt}
\begin{subfigure}[t]{0.49\textwidth}
\centering
\includegraphics[width=\linewidth]{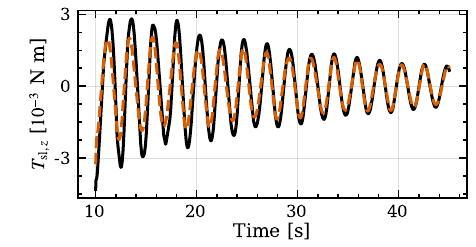}
\caption{Prescribed motion: response.}
\end{subfigure}
\hfill
\begin{subfigure}[t]{0.49\textwidth}
\centering
\includegraphics[width=\linewidth]{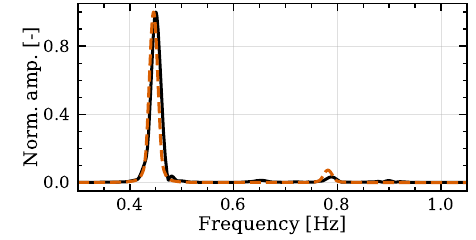}
\caption{Prescribed motion: spectrum.}
\end{subfigure}
\par\nointerlineskip\vspace{4pt}
\begin{subfigure}[t]{0.49\textwidth}
\centering
\includegraphics[width=\linewidth]{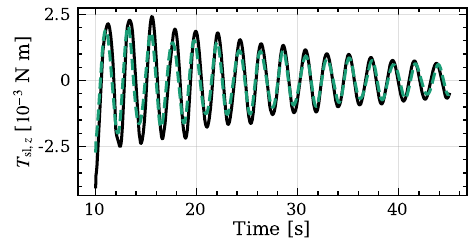}
\caption{Rigid FSI: response.}
\end{subfigure}
\hfill
\begin{subfigure}[t]{0.49\textwidth}
\centering
\includegraphics[width=\linewidth]{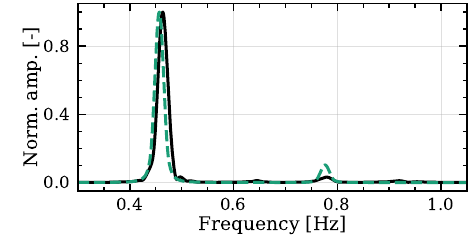}
\caption{Rigid FSI: spectrum.}
\end{subfigure}
\par\nointerlineskip\vspace{4pt}
\begin{subfigure}[t]{0.49\textwidth}
\centering
\includegraphics[width=\linewidth]{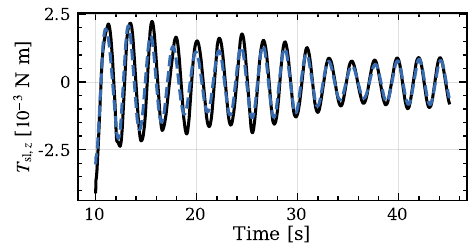}
\caption{Flexible FSI: response.}
\end{subfigure}
\hfill
\begin{subfigure}[t]{0.49\textwidth}
\centering
\includegraphics[width=\linewidth]{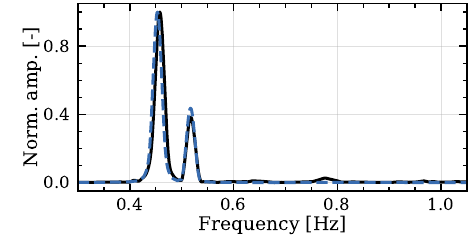}
\caption{Flexible FSI: spectrum.}
\end{subfigure}
\par\nointerlineskip\vspace{4pt}
\caption{Calibration of the two-mode physical liquid model in the prescribed-motion, rigid-FSI, and flexible-FSI architectures. In each row, the black solid trace is the DIVA reference and the coloured dashed trace is the corresponding calibrated LTI response; the right column compares their normalized spectra. Rows correspond to prescribed motion, rigid FSI, and flexible FSI; the fitted modal frequencies are respectively (0.446, 0.782), (0.440, 0.773), and (0.438, 0.799) Hz.}
\label{fig:system_aware_calibration}
\end{figure}

\begin{figure}[!htbp]
\centering
\includegraphics[width=0.49\textwidth]{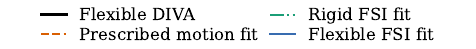}\par\nointerlineskip\vspace{1pt}
\begin{subfigure}[t]{0.49\textwidth}
\centering
\includegraphics[width=\linewidth]{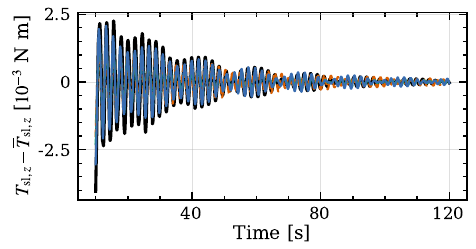}
\caption{Sloshing torque: complete replay.}
\end{subfigure}
\hfill
\begin{subfigure}[t]{0.49\textwidth}
\centering
\includegraphics[width=\linewidth]{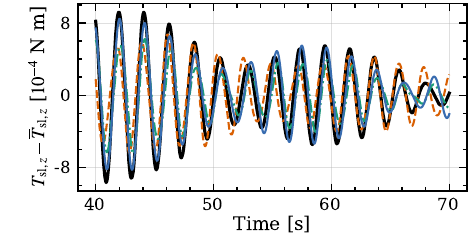}
\caption{Sloshing torque: phase detail.}
\end{subfigure}
\par\nointerlineskip\vspace{4pt}
\begin{subfigure}[t]{0.49\textwidth}
\centering
\includegraphics[width=\linewidth]{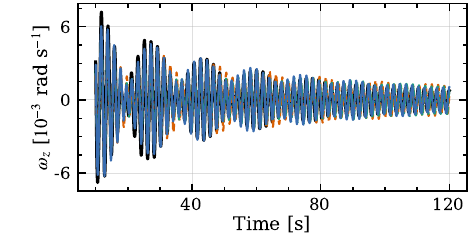}
\caption{Yaw rate: complete replay.}
\end{subfigure}
\hfill
\begin{subfigure}[t]{0.49\textwidth}
\centering
\includegraphics[width=\linewidth]{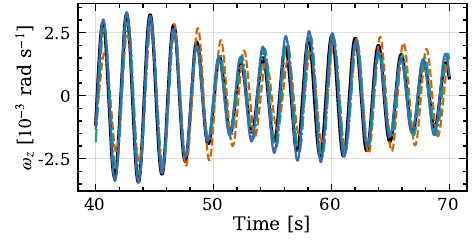}
\caption{Yaw rate: phase detail.}
\end{subfigure}
\par\nointerlineskip\vspace{4pt}
\caption{Open-loop prediction of the flexible-FSI record after inserting the PM-, rigid-FSI-, or flexible-FSI-identified liquid model into the same flexible LTI plant. The left column shows the complete post-spin response and the right column the \(40\)--\(\SI{70}{s}\) phase detail.}
\label{fig:lti_source_replay}
\end{figure}

\begin{table*}[!ht]
\centering
\caption{Common flexible-plant replay obtained with the three identified liquid models. Torque NRMSE values use mean-removed signals; the yaw-rate metric uses linear detrending.}
\label{tab:lti_source_replay}
\small
\begin{tabular}{lrrrrrr}
\toprule
Identification source &
\(\mathrm{NRMSE}_{T_{\rm sl}}\) [\%] & \(\rho_{T_{\rm sl}}\) &
\(\mathrm{NRMSE}_{T_{\rm flex}}\) [\%] & \(\rho_{T_{\rm flex}}\) &
\(\mathrm{NRMSE}_{\widetilde{\omega}}\) [\%] & \(\rho_{\widetilde{\omega}}\) \\
\midrule
Prescribed motion & 62.5 & 0.800 & 55.1 & 0.848 & 56.0 & 0.830 \\
Rigid FSI        & 47.0 & 0.929 & 39.9 & 0.922 & 40.1 & 0.916 \\
Flexible FSI     & 39.6 & 0.926 & 15.1 & 0.990 & 22.8 & 0.974 \\
\bottomrule
\end{tabular}
\end{table*}

The calibration responses are reported in \Cref{fig:system_aware_calibration}, while \Cref{fig:lti_source_replay,tab:lti_source_replay} present the corresponding replays on the common flexible LTI plant. The metrics in \Cref{tab:lti_source_replay} are evaluated over the post-spin-up interval \(10\leq t\leq\SI{120}{s}\). For a reference signal \(\widetilde{y}\) and prediction \(\widetilde{\widehat{y}}\), the normalized error is defined as \(\mathrm{NRMSE}=100\,\mathrm{RMSE}(\widetilde{y},\widetilde{\widehat{y}})/\sigma(\widetilde{y})\), and \(\rho\) is the Pearson correlation coefficient. The torque signals are mean-removed and the yaw-rate signals are linearly detrended before evaluation.
Among the three identification sources, the flexible-FSI model gives the lowest prediction errors for the sloshing torque, flexible torque, and yaw rate, while the rigid-FSI model generally provides intermediate accuracy. The benefit of flexible-FSI identification is most evident in the flexible-load and yaw-rate responses, showing that identification within the coupled flexible architecture better preserves the phase of the common-plant response. 

The common-plant replay shows that a liquid model identified from prescribed-motion data produces larger long-horizon phase errors when used in the coupled rigid--flexible system. This result indicates that the identified sloshing model is not only a property of the isolated liquid motion, but also depends on the dynamic environment in which the liquid response is observed. For the present configuration, where the liquid and flexible modes are close in frequency, identification from the flexible-FSI simulation provides the most representative reduced-order model. An equivalent coupled architecture is therefore required when the objective is to accurately predict the time-domain response of the full liquid--rigid--flexible system.

\subsection{Closed-loop CFD replay}
\label{sec:cl_campaign}

The closed-loop test uses the \(128^3\) flexible-FSI reference case and the LQG controller synthesized from the liquid model identified in the corresponding flexible simulation. The yaw-rate command is trapezoidal: the manoeuvre starts at \(t=\SI{5}{s}\), reaches \(\SI{1}{rad.s^{-1}}\) after a \(\SI{10}{s}\) ramp, remains constant for \(\SI{10}{s}\), and returns to zero through a second \(\SI{10}{s}\) ramp; the system then coasts to \(t=\SI{80}{s}\). The CFD--FSI solver supplies the measured yaw rate to the feedback loop, while the liquid and flexible reaction torques are retained as nonlinear load outputs. The purpose of the replay is not to establish controller optimality. It is to execute a controller validation step by testing the controller designed from a reduced coupled model in the nonlinear CFD--FSI plant and to quantify, signal by signal, the agreement and model-form error between the LTI design model and the full simulation.

\begin{figure}[!htbp]
\centering
\includegraphics[width=0.49\textwidth]{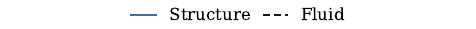}\par\nointerlineskip\vspace{1pt}
\begin{subfigure}[t]{0.49\textwidth}
\centering
\includegraphics[width=\linewidth]{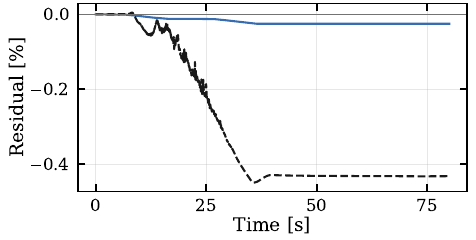}
\caption{Subsystem energy residuals.}
\end{subfigure}
\hfill
\begin{subfigure}[t]{0.49\textwidth}
\centering
\includegraphics[width=\linewidth]{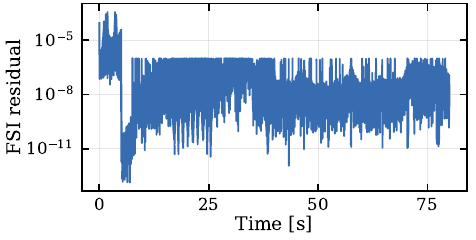}
\caption{Fluid--solid coupling residual.}
\end{subfigure}
\par\nointerlineskip\vspace{4pt}
\caption{Subsystem energy residuals and fixed-point coupling residual for the flexible-FSI LQG replay. The subsystem residuals are normalized by the maximum absolute cumulative control work.}
\label{fig:lqg_fsi_energy_and_coupling}
\end{figure}

The numerical consistency of the closed-loop replay is first assessed through the subsystem energy balances and the fixed-point coupling residual shown in \Cref{fig:lqg_fsi_energy_and_coupling}. The coupling residual remains controlled throughout the simulation, indicating that the Dirichlet--Neumann iterations consistently reconcile the motion imposed on the fluid subsystem with the reaction loads transferred to the structural subsystem. The structure and fluid residuals remain bounded over the complete replay. At the final time, the total balance residual is \(-4.70\times10^{-4}\,\si{J}\), corresponding to \(-0.457\%\) of the maximum absolute cumulative control work; the fluid and structural contributions are \(-4.45\times10^{-4}\,\si{J}\) and \(-2.57\times10^{-5}\,\si{J}\), respectively. These values serve as energy consistency check for the system and provide an independent check that the closed-loop solution is numerically consistent and the nonlinear load exchange remains numerically controlled during the closed-loop test.

\begin{figure}[p]
\centering
\includegraphics[width=\textwidth]{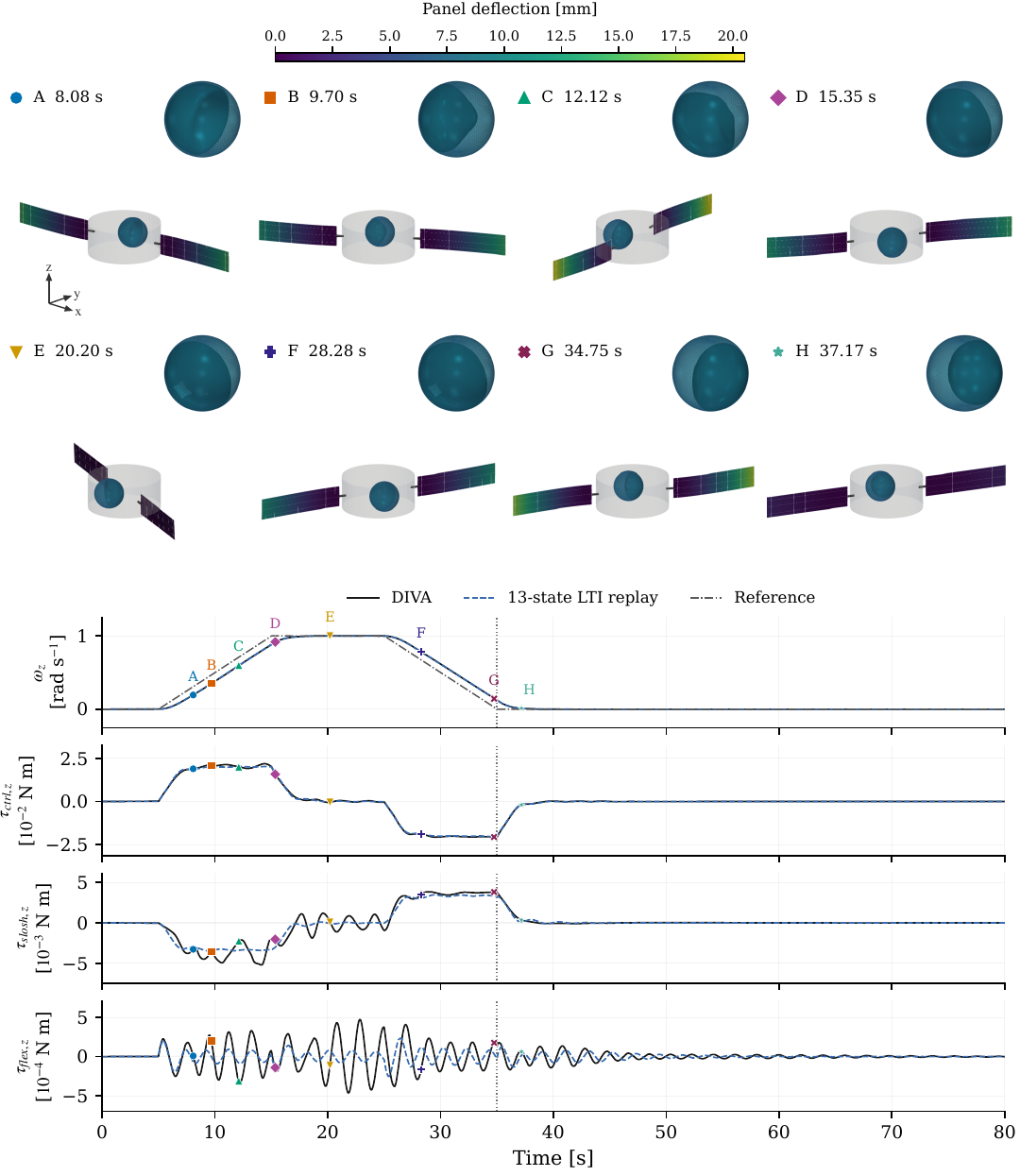}
\caption{Closed-loop flexible-FSI CFD replay with the LQG controller synthesized from the flexible-FSI liquid model. Snapshots A--H and tank zooms correspond to the marked curve instants. The tank zooms are shown in a tank-fixed reference frame, with the origin at the tank centre and viewing axes rotating with the tank, so that the liquid motion is observed relative to the wall. The full spacecraft views retain the inertial orientation; the triad below spacecraft A indicates the inertial axes. In the tank views, the darker region represents the gas bubble, while the lighter blue region represents the liquid. From top to bottom, the histories show yaw rate, control torque, sloshing torque, and flexible torque. Solid black curves show CFD--FSI; blue dashed curves show the 13-state LTI replay. The vertical marker denotes the end of the manoeuvre.}
\label{fig:lqg_fsi_closed_loop_response}
\end{figure}

The nonlinear CFD--FSI histories and the corresponding LTI replay are compared in \Cref{fig:lqg_fsi_closed_loop_response}. The attitude channel is reproduced closely: over the manoeuvre interval \(t=5\)--\SI{35}{s}, the CFD yaw-rate tracking RMSE is \(8.97\times10^{-2}\,\si{rad.s^{-1}}\), with a peak error of \(1.23\times10^{-1}\,\si{rad.s^{-1}}\), and the peak actuator torque is \(2.20\times10^{-2}\,\si{N.m}\). Over the full record, the LTI prediction has \(0.267\%\) yaw-rate NRMSE and \(3.77\%\) actuator-torque NRMSE. The agreement in these channels is expected for the selected manoeuvre. The controller bandwidth is set one decade below the first identified sloshing frequency, and the trapezoidal command therefore acts mainly in a band where the liquid and panel modes are weakly excited.
The load histories expose the limitation of the reduced liquid representation. The full-record sloshing-torque NRMSE is \(26.2\%\). The LTI model captures the dominant low-frequency attitude effect of the liquid, but it does not reproduce the observed nonlinear liquid and flexible-load histories. The discrepancy reflects differences in the resolved liquid response, including nonlinear interface motion and its interaction with the flexible structure, which are represented only approximately by the identified two-mode linear model. The data therefore establish a model-form mismatch between the two-mode surrogate and the CFD--FSI response.
This comparison is not a demonstration of general controller optimality or robustness, as only a single manoeuvre is examined. Rather, it illustrates the intended role of the coupled CFD--FSI framework: controllers synthesized from fast reduced-order models can be systematically replayed in the nonlinear plant to evaluate the coupled closed-loop dynamics and quantify model-form errors. Integrating the 13-state LTI model required a median of \(\SI{3.59}{s}\) over three runs on an Intel Core i7-11850H processor at \(\SI{2.50}{GHz}\). The corresponding nonlinear $128^3$ CFD--FSI replay required \(\SI{17.5}{h}\) on 128 MPI ranks of the CALMIP Olympe system, equivalent to approximately \(2.24\times10^3\) MPI-rank hours, to simulate \(\SI{80}{s}\) of physical time. Although such computational cost precludes direct on-board implementation or iterative controller-design sweeps, the high-fidelity coupled solver remains an indispensable testbed for evaluating model-form errors and assessing closed-loop behaviour in the presence of nonlinear sloshing.

\section{Conclusions}
\label{sec:conclusions}

This paper developed and assessed a partitioned DNS–FSI framework for the simulation of coupled liquid–rigid–flexible spacecraft dynamics under microgravity conditions. The formulation combines the FLUIDICS-validated incompressible two-phase flow solver implemented in DIVA with rigid-body attitude dynamics and a rotating flexible-panel model, and enforces two-way consistency between the tank angular kinematics and the hydrodynamic loads through a strongly coupled Dirichlet–Neumann fixed-point algorithm.
The study pursued three complementary objectives: to assess the implementation of the coupled load–motion exchange, to determine how rigid-body and flexible-structure feedback modify the sloshing response, and to demonstrate how the nonlinear framework can be used to evaluate a controller synthesized from a reduced coupled model.

The coupling implementation was first assessed against the free-decay experiment of Peterson et al.~\cite{peterson1989nonlinear} The coupled calculation reproduced the dominant oscillation frequency and maintained the phase of the measured tank displacement over the reported comparison interval. The simulation also recovered the reduction in system frequency relative to the frozen-liquid response, consistent with the effective dynamic compliance introduced by relative liquid motion. This comparison supports the implementation of the partitioned exchange for a coupled sloshing problem.

A closed-system mechanical energy balance was derived and used as an a posteriori numerical verification of the coupled simulations. Spatial refinement consistently reduced the final deficit. This trend indicates that the dominant contribution is associated with the spatial representation of the liquid domain, the interface and immersed boundary, and the fluid loads entering the coupled energy exchange.

The comparison between prescribed-motion, rigid open-loop, and flexible open-loop simulations showed that two-way coupling modifies the liquid response even when the global attitude histories remain relatively close. The prescribed-motion calculation does not account for the feedback of the liquid loads on the tank motion, while the rigid and flexible calculations allow the spacecraft dynamics to alter the excitation experienced by the liquid. The spectral analysis further showed that structural flexibility can modify the sloshing response when structural and liquid frequencies are close. In the present configuration, the flexible simulation exhibited a different persistence and spectral composition of the liquid torque, including a component shared with the panel response that was absent from the prescribed-motion and rigid cases. These results indicate that the liquid and flexible structure cannot always be treated as dynamically independent subsystems: when their characteristic frequencies overlap, their interaction can alter the observed sloshing dynamics and the loads transmitted to the spacecraft.
This dependence also affected reduced-order identification. Two-mode liquid surrogates identified independently from the prescribed-motion, rigid-FSI, and flexible-FSI responses produced different long-time predictions when inserted into the same rigid–flexible LTI plant. The model identified from the flexible-FSI record gave the most representative prediction for the corresponding coupled system, whereas identification from prescribed-motion data produced larger phase discrepancies. Thus, the identified liquid model was not solely a property of the tank and liquid configuration; within the adopted identification procedure, it also reflected the dynamic environment and excitation from which the response was obtained. From a control perspective, this implies that a reduced sloshing model calibrated from prescribed kinematics may not retain the phase and load information required to represent the same liquid once it is embedded in a coupled flexible spacecraft.

The final closed-loop replay demonstrated the control-assessment role of the framework. An LQG controller synthesized from the reduced model identified from the flexible simulation was applied without modification to the nonlinear $128^3$ CFD–FSI plant. For the considered trapezoidal yaw-rate command, the nonlinear system tracked accurately the prescribed manoeuvre. The LTI and nonlinear histories remained close in the principal closed-loop variables, with NRMSE values of $0.267\%$ for yaw rate and $3.77\%$ for actuator torque. However, the sloshing-torque NRMSE reached $26.2\%$ . The reduced model therefore reproduced the low-frequency attitude and control response for this manoeuvre but not the detailed nonlinear load history. 
The closed-loop study therefore highlights the complementary role of reduced-order and high-fidelity models. Reduced models remain appropriate for controller synthesis and rapid system-level analysis, but agreement in the commanded attitude response does not guarantee an equally accurate prediction of the internal liquid and flexible loads. This limitation becomes more important when the controller bandwidth approaches the dominant structural and sloshing frequencies, since the manoeuvre can then excite coupled modes for which linearized liquid models are less representative. In such regimes, the proposed CFD–FSI framework provides a high-fidelity nonlinear environment for investigating the coupled response, quantifying model-form discrepancies, and validating control strategies under conditions that are not adequately described by the synthesis model.

\section*{Acknowledgements}

This work was supported by the Open Space Innovation Platform (OSIP) of the European Space Agency (ESA), Contract No. 4000143096/23/NL/MGu/my; the von Karman Institute, Contract No. ARD2407; and the Chair SaCLaB2, resulting from the partnership between Airbus Defence and Space, ArianeGroup, and ISAE-SUPAERO, which co-funds the PhD thesis of Umberto Zucchelli. Miguel Alfonso Mendez is funded by the European Research Council (ERC) under the European Union's Horizon Europe programme (RE-TWIST project, grant agreement No 101165479). The views expressed are those of the authors and do not necessarily reflect those of the European Union or the ERC.
This work was granted access to the high-performance computing resources of CALMIP under project allocation p23052.

\section*{Declaration of competing interests}

The authors declare that they have no known competing financial interests or personal relationships that could have appeared to influence the work reported in this paper.

\section*{CRediT authorship contribution statement}

\textbf{Umberto Zucchelli}: Conceptualization, Methodology, Software, Validation, Formal analysis, Investigation, Data curation, Writing - original draft, Visualization.
\textbf{Miguel Alfonso Mendez}: Conceptualization, Methodology, Formal analysis, Supervision, Writing - review \& editing.
\textbf{Annafederica Urbano}: Methodology, Software, Formal analysis, Writing - review \& editing.
\textbf{Sebastien Vincent-Bonnieu}: Conceptualization, Resources.
\textbf{Piotr Wenderski}: Conceptualization, Resources.
\textbf{Francesco Sanfedino}: Conceptualization, Methodology, Formal analysis, Supervision, Project administration, Writing - review \& editing.

\bibliographystyle{unsrtnat}
\bibliography{references}

@techreport{Abramson1966,
  author = {Abramson, H. N.},
  title = {The dynamic behavior of liquids in moving containers, with applications to space vehicle technology},
  institution = {NASA},
  type = {NASA Special Publication},
  year = {1966},
  number = {SP-106},
  url = {https://ntrs.nasa.gov/citations/19670006555}
}

@book{Dodge2000,
  author = {Dodge, F. T.},
  title = {The new ``Dynamic Behavior of Liquids in Moving Containers''},
  publisher = {Southwest Research Institute},
  year = {2000},
  address = {San Antonio, TX},
  url = {https://www2.swri.org/www2/fluid-slosh/slosh-dynamics.htm}
}

@article{Hoskoti2023,
  author = {Hoskoti, L. and Gupta, S. S. and Sucheendran, M. M.},
  title = {Modeling of geometrical stiffening in a rotating blade---A review},
  journal = {J. Sound Vib.},
  year = {2023},
  volume = {548},
  pages = {117526},
  doi = {10.1016/j.jsv.2022.117526}
}

@article{Rodrigues2024,
  author = {Rodrigues, R. and Alazard, D. and Sanfedino, F. and Mauriello, T. and Iannelli, P.},
  title = {Modeling and analysis of a flexible spinning {Euler--Bernoulli} beam with centrifugal stiffening and softening: A linear fractional representation approach with application to spinning spacecraft},
  journal = {Appl. Math. Model.},
  volume = {137},
  pages = {115699},
  year = {2025},
  doi = {10.1016/j.apm.2024.115699}
}

@article{simonini_2024,
  author = {Simonini, A. and Dreyer, M. and Urbano, A. and Sanfedino, F. and Himeno, T. and Behruzi, P. and Avila, M. and Pinho, J. and Peveroni, L. and Gouriet, J.-B.},
  title = {Cryogenic propellant management in space: open challenges and perspectives},
  journal = {npj Microgravity},
  volume = {10},
  pages = {34},
  year = {2024},
  doi = {10.1038/s41526-024-00377-5},
  number = {1}
}

@article{Gasbarri2016,
  author = {Posani, M. and Pontani, M. and Gasbarri, P.},
  title = {Nonlinear Slewing Control of a Large Flexible Spacecraft Using Reaction Wheels},
  journal = {Aerospace},
  year = {2022},
  volume = {9},
  pages = {244},
  doi = {10.3390/aerospace9050244},
  number = {5}
}

@article{Dalmon2019,
  author = {Dalmon, A. and Lepilliez, M. and Tanguy, S. and Alis, R. and Popescu, E. R. and Roumigui{\'e}, R. and Miquel, T. and Busset, B. and Bavestrello, H. and Mignot, J.},
  title = {Comparison between the {FLUIDICS} experiment and direct numerical simulations of fluid sloshing in spherical tanks under microgravity conditions},
  journal = {Microgravity Sci. Technol.},
  volume = {31},
  number = {1},
  pages = {123--138},
  year = {2019},
  doi = {10.1007/s12217-019-9675-4}
}

@inproceedings{VallesSanchez2025,
  author = {Zucchelli, U. and Valles S{\'a}nchez, I. and Sanfedino, F.},
  title = {Modeling of spinning plates: geometric stiffening and modal approximation for {GNC} applications},
  booktitle = {Proceedings of the {CEAS--AIDAA} 2025 Conference},
  year = {2025},
  note = {Paper 426}
}

@article{Tanguy2005,
  author = {Tanguy, S. and Berlemont, A.},
  title = {Application of a level set method for simulation of droplet collisions},
  journal = {Int. J. Multiph. Flow},
  volume = {31},
  pages = {1015--1035},
  year = {2005},
  doi = {10.1016/j.ijmultiphaseflow.2005.05.010}
}

@article{Lalanne2015,
  author = {Lalanne, B. and Abi Chebel, N. and Vejra{\v z}ka, J. and Tanguy, S. and Masbernat, O. and Risso, F.},
  title = {Non-linear shape oscillations of rising drops and bubbles: experiments and simulations},
  journal = {Phys. Fluids},
  volume = {27},
  pages = {123305},
  year = {2015},
  doi = {10.1063/1.4936980}
}

@article{RuedaVillegas2017,
  author = {Rueda Villegas, L. and Tanguy, S. and Castanet, G. and Caballina, O. and Lemoine, F.},
  title = {Direct numerical simulation of the impact of a droplet onto a hot surface above the {Leidenfrost} temperature},
  journal = {Int. J. Heat Mass Transf.},
  volume = {104},
  pages = {1090--1109},
  year = {2017},
  doi = {10.1016/j.ijheatmasstransfer.2016.08.105}
}

@article{Urbano2018,
  author = {Urbano, A. and Tanguy, S. and Huber, G. and Colin, C.},
  title = {Direct numerical simulation of nucleate boiling in micro-layer regime},
  journal = {Int. J. Heat Mass Transf.},
  volume = {123},
  pages = {1128--1137},
  year = {2018},
  doi = {10.1016/j.ijheatmasstransfer.2018.02.104}
}

@article{Sussman1994,
  author = {Sussman, M. and Smereka, P. and Osher, S.},
  title = {A level set approach for computing solutions to incompressible two-phase flow},
  journal = {J. Comput. Phys.},
  volume = {114},
  pages = {146--159},
  year = {1994},
  doi = {10.1006/jcph.1994.1155}
}

@article{Jiang1996,
  author = {Jiang, G.-S. and Shu, C.-W.},
  title = {Efficient implementation of weighted {ENO} schemes},
  journal = {J. Comput. Phys.},
  volume = {126},
  pages = {202--228},
  year = {1996},
  doi = {10.1006/jcph.1996.0130}
}

@article{BorgesEtAl2008,
  author = {Borges, R. and Carmona, M. and Costa, B. and Don, W. S.},
  title = {An improved weighted essentially non-oscillatory scheme for hyperbolic conservation laws},
  journal = {J. Comput. Phys.},
  volume = {227},
  pages = {3191--3211},
  year = {2008},
  doi = {10.1016/j.jcp.2007.11.038}
}

@article{Aslam2004,
  author = {Aslam, T. D.},
  title = {A partial differential equation approach to multidimensional extrapolation},
  journal = {J. Comput. Phys.},
  volume = {193},
  pages = {349--355},
  year = {2004},
  doi = {10.1016/j.jcp.2003.08.001}
}

@article{Lepilliez2016,
  author = {Lepilliez, M. and Popescu, E. R. and Gibou, F. and Tanguy, S.},
  title = {On two-phase flow solvers in irregular domains with contact line},
  journal = {J. Comput. Phys.},
  volume = {321},
  pages = {1217--1251},
  year = {2016},
  doi = {10.1016/j.jcp.2016.06.013}
}

@article{Chorin1967,
  author = {Chorin, A. J.},
  title = {A numerical method for solving incompressible viscous flow problems},
  journal = {J. Comput. Phys.},
  volume = {2},
  pages = {12--26},
  year = {1967},
  doi = {10.1016/0021-9991(67)90037-x}
}

@article{Shu1988,
  author = {Shu, C.-W. and Osher, S.},
  title = {Efficient implementation of essentially non-oscillatory shock-capturing schemes},
  journal = {J. Comput. Phys.},
  volume = {77},
  pages = {439--471},
  year = {1988},
  doi = {10.1016/0021-9991(88)90177-5}
}

@article{Liu2000,
  author = {Liu, X.-D. and Fedkiw, R. and Kang, M.},
  title = {A boundary condition capturing method for {Poisson}'s equation on irregular domains},
  journal = {J. Comput. Phys.},
  volume = {160},
  pages = {151--178},
  year = {2000},
  doi = {10.1006/jcph.2000.6444}
}

@article{Fedkiw1999,
  author = {Fedkiw, R. and Aslam, T. and Merriman, B. and Osher, S.},
  title = {A non-oscillatory {Eulerian} approach to interfaces in multimaterial flows (the ghost fluid method)},
  journal = {J. Comput. Phys.},
  volume = {152},
  pages = {457--492},
  year = {1999},
  doi = {10.1006/jcph.1999.6236}
}

@article{Dendy1982,
  author = {Dendy, J. E.},
  title = {Black box multigrid},
  journal = {J. Comput. Phys.},
  volume = {48},
  pages = {366--386},
  year = {1982},
  doi = {10.1016/0021-9991(82)90057-2}
}

@article{MacLachlan2008,
  author = {MacLachlan, S. P. and Tang, J. M. and Vuik, C.},
  title = {Fast and robust solvers for pressure-correction in bubbly flow problems},
  journal = {J. Comput. Phys.},
  volume = {227},
  number = {23},
  pages = {9742--9761},
  year = {2008},
  doi = {10.1016/j.jcp.2008.07.022}
}

@article{Gibou2002,
  author = {Gibou, F. and Fedkiw, R. and Cheng, L. T. and Kang, M.},
  title = {A second-order-accurate symmetric discretization of the {Poisson} equation on irregular domains},
  journal = {J. Comput. Phys.},
  volume = {176},
  pages = {205--227},
  year = {2002},
  doi = {10.1006/jcph.2001.6977}
}

@article{Ng2009,
  author = {Ng, Y. T. and Min, C. and Gibou, F.},
  title = {An efficient fluid--solid coupling algorithm for single-phase flows},
  journal = {J. Comput. Phys.},
  volume = {228},
  pages = {8807--8829},
  year = {2009},
  doi = {10.1016/j.jcp.2009.08.032}
}

@article{Kuettler2008,
  author = {K{\"u}ttler, U. and Wall, W. A.},
  title = {Fixed-point fluid--structure interaction solvers with dynamic relaxation},
  journal = {Comput. Mech.},
  volume = {43},
  pages = {61--72},
  year = {2008},
  doi = {10.1007/s00466-008-0255-5}
}

@article{Guermond2006,
  author = {Guermond, J.-L. and Minev, P. and Shen, J.},
  title = {An overview of projection methods for incompressible flows},
  journal = {Comput. Methods Appl. Mech. Engrg.},
  volume = {195},
  pages = {6011--6045},
  year = {2006},
  doi = {10.1016/j.cma.2005.10.010}
}

@inproceedings{Mignot2017,
  author = {Mignot, J. and Pierre, R. and Berhanu, M. and Busset, B. and Roumigui{\'e}, R. and Bavestrello, H. and Bonfanti, S. and Miquel, T. and Oro Marot, L. and Llodra-Perez, A.},
  title = {Fluid dynamic in space experiment},
  booktitle = {Proceedings of the 68th International Astronautical Congress ({IAC})},
  address = {Adelaide, Australia},
  year = {2017},
  note = {Paper IAC-17-A2.6.2},
  url = {https://labo.msc.u-paris.fr/~berhanu/papers/IAC-2017-final-Fluidics.pdf}
}

@article{Farhat1998,
  author = {Farhat, C. and Lesoinne, M. and Le Tallec, P.},
  title = {Load and motion transfer algorithms for fluid--structure interaction problems with non-matching discrete interfaces: momentum and energy conservation, optimal discretization and application to aeroelasticity},
  journal = {Comput. Methods Appl. Mech. Engrg.},
  volume = {157},
  pages = {95--114},
  year = {1998},
  doi = {10.1016/s0045-7825(97)00216-8}
}

@article{Bungartz2016,
  author = {Bungartz, H.-J. and Lindner, F. and Gatzhammer, B. and Mehl, M. and Scheufele, K. and Shukaev, A. and Uekermann, B.},
  title = {{preCICE} -- a fully parallel library for multi-physics surface coupling},
  journal = {Comput. Fluids},
  volume = {141},
  pages = {250--258},
  year = {2016},
  doi = {10.1016/j.compfluid.2016.04.003}
}

@article{Ye1999,
  author = {Ye, T. and Mittal, R. and Udaykumar, H. S. and Shyy, W.},
  title = {An accurate {Cartesian} grid method for viscous incompressible flows with complex immersed boundaries},
  journal = {J. Comput. Phys.},
  volume = {156},
  pages = {209--240},
  year = {1999},
  doi = {10.1006/jcph.1999.6356}
}

@article{TsengFerziger2003,
  author = {Tseng, Y.-H. and Ferziger, J. H.},
  title = {A ghost-cell immersed boundary method for flow in complex geometry},
  journal = {J. Comput. Phys.},
  volume = {192},
  pages = {593--623},
  year = {2003},
  doi = {10.1016/j.jcp.2003.07.024}
}

@article{TairaColonius2007,
  author = {Taira, K. and Colonius, T.},
  title = {The immersed boundary method: a projection approach},
  journal = {J. Comput. Phys.},
  volume = {225},
  pages = {2118--2137},
  year = {2007},
  doi = {10.1016/j.jcp.2007.03.005}
}

@phdthesis{DalmonThesis2019,
  author = {Dalmon, Alexis},
  title = {Simulation num{\'e}rique du ballottement d'ergol et mod{\'e}lisation de l'interaction fluides-membrane dans un r{\'e}servoir de satellite},
  school = {Universit{\'e} de Toulouse},
  year = {2018},
  url = {https://doctorat.univ-toulouse.fr/as/ed/cv.pl?mat=71174&site=EDT}
}

@article{HarlowWelch1965,
  author = {Harlow, F. H. and Welch, J. E.},
  title = {Numerical calculation of time-dependent viscous incompressible flow of fluid with free surface},
  journal = {Phys. Fluids},
  volume = {8},
  number = {12},
  pages = {2182--2189},
  year = {1965},
  doi = {10.1063/1.1761178}
}

@article{Degroote2010,
  author = {Degroote, J. and Souto-Iglesias, A. and Van Paepegem, W. and Annerel, S. and Bruggeman, P. and Vierendeels, J.},
  title = {Partitioned simulation of the interaction between an elastic structure and free surface flow},
  journal = {Comput. Methods Appl. Mech. Engrg.},
  volume = {199},
  number = {33--36},
  pages = {2085--2098},
  year = {2010},
  doi = {10.1016/j.cma.2010.02.019}
}

@article{MittalIaccarino2005,
  author = {Mittal, R. and Iaccarino, G.},
  title = {Immersed boundary methods},
  journal = {Annu. Rev. Fluid Mech.},
  volume = {37},
  pages = {239--261},
  year = {2005},
  doi = {10.1146/annurev.fluid.37.061903.175743}
}

@article{Shu2009WENO,
  author = {Shu, C.-W.},
  title = {High order weighted essentially nonoscillatory schemes for convection dominated problems},
  journal = {SIAM Rev.},
  volume = {51},
  number = {1},
  pages = {82--126},
  year = {2009},
  doi = {10.1137/070679065}
}

@book{Tryggvason2011,
  author = {Tryggvason, G. and Scardovelli, R. and Zaleski, S.},
  title = {Direct Numerical Simulations of Gas--Liquid Multiphase Flows},
  publisher = {Cambridge University Press},
  year = {2011}
}

@article{SchotteOhayon2009,
  author = {Schott{\'e}, J.-S. and Ohayon, R.},
  title = {Various modelling levels to represent internal liquid behaviour in the vibration analysis of complex structures},
  journal = {Comput. Methods Appl. Mech. Engrg.},
  volume = {198},
  pages = {1913--1925},
  year = {2009},
  doi = {10.1016/j.cma.2008.12.016}
}

@article{ChoLee2004,
  author = {Cho, J. R. and Lee, H. W.},
  title = {Numerical study on liquid sloshing in baffled tank by nonlinear finite element method},
  journal = {Comput. Methods Appl. Mech. Engrg.},
  volume = {193},
  pages = {2581--2598},
  year = {2004},
  doi = {10.1016/j.cma.2004.01.009}
}

@article{IdelsohnOnateDelPinCalvo2006,
  author = {Idelsohn, S. R. and O{\~n}ate, E. and Del Pin, F. and Calvo, N.},
  title = {Fluid--structure interaction using the particle finite element method},
  journal = {Comput. Methods Appl. Mech. Engrg.},
  volume = {195},
  pages = {2100--2123},
  year = {2006},
  doi = {10.1016/j.cma.2005.02.026}
}

@article{PanCaoLi2020,
  author = {Pan, K. and Cao, D. and Li, J.},
  title = {Absolute nodal coordinate particle finite element to the free-surface flow problems combined with multibody algorithms},
  journal = {Comput. Methods Appl. Mech. Engrg.},
  volume = {372},
  pages = {113378},
  year = {2020},
  doi = {10.1016/j.cma.2020.113378}
}

@article{PillotonSunZhangColagrossi2024,
  author = {Pilloton, C. and Sun, P. N. and Zhang, X. and Colagrossi, A.},
  title = {Volume conservation issue within {SPH} models for long-time simulations of violent free-surface flows},
  journal = {Comput. Methods Appl. Mech. Engrg.},
  volume = {419},
  pages = {116640},
  year = {2024},
  doi = {10.1016/j.cma.2023.116640}
}

@article{YuYue2023,
  author = {Yu, J. and Yue, B.},
  title = {Study on the coupled dynamics of rigid-liquid-flexible spacecraft by using isogeometric analysis for liquid sloshing},
  journal = {Appl. Math. Model.},
  volume = {113},
  pages = {88--108},
  year = {2023},
  doi = {10.1016/j.apm.2022.08.026}
}

@article{MonteleoneBorinoNapoliBurriesci2022,
  author = {Monteleone, A. and Borino, G. and Napoli, E. and Burriesci, G.},
  title = {Fluid--structure interaction approach with smoothed particle hydrodynamics and particle--spring systems},
  journal = {Comput. Methods Appl. Mech. Engrg.},
  volume = {392},
  pages = {114728},
  year = {2022},
  doi = {10.1016/j.cma.2022.114728}
}

@article{SunColagrossiMarroneAntuonoZhang2019,
  author = {Sun, P. N. and Colagrossi, A. and Marrone, S. and Antuono, M. and Zhang, A.-M.},
  title = {A consistent approach to particle shifting in the {$\delta$}-plus-{SPH} model},
  journal = {Comput. Methods Appl. Mech. Engrg.},
  volume = {348},
  pages = {912--934},
  year = {2019},
  doi = {10.1016/j.cma.2019.01.045}
}

@article{GaoFu2023,
  author = {Gao, T. and Fu, L.},
  title = {A new particle shifting technique for {SPH} methods based on {Voronoi} diagram and volume compensation},
  journal = {Comput. Methods Appl. Mech. Engrg.},
  volume = {404},
  pages = {115788},
  year = {2023},
  doi = {10.1016/j.cma.2022.115788}
}

@article{VergnaudOgerLeTouzeDeLeffeChiron2022,
  author = {Vergnaud, A. and Oger, G. and Le Touz{\'e}, D. and DeLeffe, M. and Chiron, L.},
  title = {{C-CSF}: Accurate, robust and efficient surface tension and contact angle models for single-phase flows using {SPH}},
  journal = {Comput. Methods Appl. Mech. Engrg.},
  volume = {389},
  pages = {114292},
  year = {2022},
  doi = {10.1016/j.cma.2021.114292}
}

@article{peterson1989nonlinear,
  author  = {Peterson, L. D. and Crawley, E. F. and Hansman, R. J.},
  title   = {Nonlinear Fluid Slosh Coupled to the Dynamics of a Spacecraft},
  journal = {AIAA J.},
  year    = {1989},
  volume  = {27},
  number  = {9},
  pages   = {1230--1240},
  doi     = {10.2514/3.10250}
}

@article{capolupo2025equivalent,
  title         = {Equivalent Mechanical Models for Sloshing},
  author        = {Capolupo, Francesco},
  journal       = {arXiv preprint arXiv:2511.10172},
  year          = {2025},
  doi           = {10.48550/arXiv.2511.10172}
}

@book{ibrahim2005liquid,
  title={Liquid Sloshing Dynamics: Theory and Applications},
  author={Ibrahim, Raouf A.},
  year={2005},
  publisher={Cambridge University Press},
  address={Cambridge, UK},
  doi={10.1017/CBO9780511536656}
}

@article{Sussman2007,
 author = {Sussman, M. and Smith, K. M. and Hussaini, M. Y. and Ohta, M. and Zhi-Wei, R.},
 title = {A sharp interface method for incompressible two-phase flows},
 journal = {Journal of Computational Physics},
 volume = {221}, number = {2}, pages = {469--505}, year = {2007},
 doi = {10.1016/j.jcp.2006.06.020}
}

\newpage

\appendix
\section{Control-oriented state-space matrices}
\label{app:control_lti_matrices}

This appendix gives the explicit matrices used to pass from the second-order reduced model in \Cref{eq:control_second_order_yaw_flex} to the first-order LTI plant in \Cref{eq:control_lti_state}. Define
\begin{equation}
\mat{M}_g=
\begin{bmatrix}
I_{\mathrm{base}} & \vect{p}_z^T \\
\vect{p}_z & \mat{M}_f
\end{bmatrix},
\qquad
\mat{H}=\mat{M}_g^{-1}=
\begin{bmatrix}
h_{00} & \vect{h}_{0f}^{T} \\
\vect{h}_{f0} & \mat{H}_{ff}
\end{bmatrix} .
\label{eq:appendix_mass_inverse_partition}
\end{equation}
The scalar \(h_{00}\), the vectors \(\vect{h}_{0f}\) and \(\vect{h}_{f0}\), and the matrix \(\mat{H}_{ff}\) are the yaw--flexible blocks of the inverse generalized mass matrix. With
\begin{equation}
\vect{R}_s=\begin{bmatrix}R_1 & \cdots & R_{N_s}\end{bmatrix}^{T},
\quad
\mat{\Omega}_s^2=\operatorname{diag}\left(\omega_{s,1}^2,\ldots,\omega_{s,N_s}^2\right),
\quad
\mat{D}_s=\operatorname{diag}\left(2\zeta_{s,1}\omega_{s,1},\ldots,2\zeta_{s,N_s}\omega_{s,N_s}\right),
\label{eq:appendix_sloshing_diagonal_matrices}
\end{equation}
and with \(\vect{1}_s\in\mathbb{R}^{N_s}\) a vector of ones, the first-order equations are
\begin{align}
\dot{\omega}_z
&=h_{00}u+h_{00}\vect{R}_s^T\vect{\xi}_s
-\vect{h}_{0f}^{T}\mat{K}_f\vect{\eta}_f
-\vect{h}_{0f}^{T}\mat{D}_f\vect{\nu}_f, \\
\dot{\vect{\xi}}_s
&=\vect{\nu}_s, \\
\dot{\vect{\nu}}_s
&=\vect{1}_s\dot{\omega}_z
-\mat{\Omega}_s^2\vect{\xi}_s
-\mat{D}_s\vect{\nu}_s, \\
\dot{\vect{\eta}}_f
&=\vect{\nu}_f, \\
\dot{\vect{\nu}}_f
&=\vect{h}_{f0}u+\vect{h}_{f0}\vect{R}_s^T\vect{\xi}_s
-\mat{H}_{ff}\mat{K}_f\vect{\eta}_f
-\mat{H}_{ff}\mat{D}_f\vect{\nu}_f .
\label{eq:appendix_first_order_rows}
\end{align}
For the state ordering of \Cref{eq:lti_state_vector}, these equations give
\begin{equation}
\mat{A}=\begin{bmatrix}
0 & h_{00}\vect{R}_s^T & \vect{0}_{1\times N_s} & -\vect{h}_{0f}^{T}\mat{K}_f & -\vect{h}_{0f}^{T}\mat{D}_f \\
\vect{0}_{N_s\times 1} & \mat{0}_{N_s\times N_s} & \mat{I}_{N_s} & \mat{0}_{N_s\times N_f} & \mat{0}_{N_s\times N_f} \\
\vect{0}_{N_s\times 1} & \vect{1}_s h_{00}\vect{R}_s^T-\mat{\Omega}_s^2 & -\mat{D}_s & -\vect{1}_s\vect{h}_{0f}^{T}\mat{K}_f & -\vect{1}_s\vect{h}_{0f}^{T}\mat{D}_f \\
\vect{0}_{N_f\times 1} & \mat{0}_{N_f\times N_s} & \mat{0}_{N_f\times N_s} & \mat{0}_{N_f\times N_f} & \mat{I}_{N_f} \\
\vect{0}_{N_f\times 1} & \vect{h}_{f0}\vect{R}_s^T & \mat{0}_{N_f\times N_s} & -\mat{H}_{ff}\mat{K}_f & -\mat{H}_{ff}\mat{D}_f
\end{bmatrix},
\label{eq:appendix_A_matrix}
\end{equation}
\begin{equation}
\vect{B}=\begin{bmatrix}
h_{00} \\
\vect{0}_{N_s} \\
h_{00}\vect{1}_s \\
\vect{0}_{N_f} \\
\vect{h}_{f0}
\end{bmatrix} .
\label{eq:appendix_B_matrix}
\end{equation}
The two outputs used in the control study are the measured angular rate and the reconstructed sloshing torque. Hence,
\begin{equation}
\mat{C}=\begin{bmatrix}
1 & \vect{0}_{1\times N_s} & \vect{0}_{1\times N_s} & \vect{0}_{1\times N_f} & \vect{0}_{1\times N_f} \\
0 & \vect{R}_s^T & \vect{0}_{1\times N_s} & \vect{0}_{1\times N_f} & \vect{0}_{1\times N_f}
\end{bmatrix} .
\label{eq:appendix_C_matrix}
\end{equation}

\end{document}